\documentclass[aps,pre,eqsecnum,showpacs,draft]{revtex4} 
\usepackage{amsmath,bm}

\begin{document}

\draft 
\title{Semiclassical quantization of bound and quasi-stationary
states beyond the adiabatic approximation.}
 
\author{V.A.Benderskii} 
\affiliation {Institute of Problems of Chemical Physics, RAS \\ 142432 Moscow
Region, Chernogolovka, Russia} 
\affiliation{Laue-Langevin Institute, F-38042,
Grenoble, France} 
 
\author{E.V.Vetoshkin} 
\affiliation {Institute of Problems of Chemical Physics, RAS \\ 142432 Moscow
Region, Chernogolovka, Russia} 
\author{E. I. Kats} \affiliation{Laue-Langevin Institute, F-38042,
Grenoble, France} 
\affiliation{L. D. Landau Institute for Theoretical Physics, RAS, Moscow, Russia}
 
\date{\today}

\begin{abstract}
We examine one important (and overlooked in all previous investigations)
aspect of well - known crossing
diabatic potentials or Landau - Zener (LZ) problem. We derive 
the semiclassical quantization rules for the crossing 
diabatic potentials with localized initial and localized or delocalized final
states, in the intermediate energy region, when all four adiabatic states
are coupled and should be taken into account. 
We found all needed connection matrices and 
present the following new analytical results:
(i) in the tunneling region the splittings of vibrational levels
are represented as a product of the
splitting in the lower adiabatic potential and the non-trivial function depending on the
Massey parameter; (ii) in the over-barrier region we find specific resonances between the levels
in the lower and in the upper adiabatic potentials and in that condition
independent quantizations rules are not correct; 
(iii) for the delocalized final states (decay lower adiabatic potential)
we describe quasi - stationary states and calculate the decay rate as a function of
the adiabatic coupling; 
(iv) for the intermediate
energy regions we calculate the energy level quantization, which can be brought into
a compact form by using either adiabatic or diabatic basis set (in contrast to the previous results
found in the Landau diabatic basis).
Applications of the results may concern the various systems, e.g.,
molecules undergoing conversion of electronic states, radiationless transitions, or
isomerization reactions.

\end{abstract}

\pacs{31.50.Gh, 05.45.-a, 72.10.-d}
\maketitle
\section{Introduction.}
\label{I}

There has been great progress in the theory of crossing potentials
during the last seven decades (see e.g. the references in both research and textbook literature
\cite{LL65} - \cite{SK02}). 
Surprisingly, seemingly a simple but a basic question how well known
semiclassical quantization rules should be modified for this particular
situation (crossing diabatic potentials with bound (i.e. localized)
initial and localized
or delocalized final
states) to the best of our knowledge are still unanswered, 
(at least a complete and unifying description of the quantization
for a general case is still not available and a number
of other questions remain to be clarified). 

Partially it is related to the fact that unfortunately experimental data
in this field are still scarce and not very accurate.
However, the situation is now changing. Experimental techniques 
(e.g., the increasing precision of experimental tests in the femtosecond
laser pulse range enables to excite well defined molecular states and to study their evolution in time
using the second probing laser beam \cite{ZE94}, 
\cite{GS92}, 
\cite{DH98}) have
progressed to the point where molecular tunneling splitting dependence on energy
can be measured in well controlled conditions with a very high accuracy.
It would therefore seem appropriate at this time
to take a fresh look at the theory, which is the purpose
of the present article.
Note also that these questions
are not only of interest in their own intellectual right.
Recent experimental and theoretical advances \cite{BO03}, \cite{MC03}
in particular are beginning to yield a coherent understanding of
several phenomena that, far from requiring minor corrections to the
standard adiabatic treatment of the problem.
Physically such kind of situations can occur as a result
of non-adiabatic interactions of different electronic states
forming in crossing one-well diabatic vibrational potentials.
Adiabatic coupling removes diabatic level crossing, and the diabatic
levels are replaced by the adiabatic ones (see Fig.1 illustrating this
phenomenon). In the case of a large adiabatic splitting (see precise
criteria below) one can restrict oneself to the only lower adiabatic potential
(symmetric or asymmetric double-well, or decay potential 
for the systems under consideration) and neglect
any influence of the upper adiabatic potential (parabolic one-well for our case).
However, in a general case of arbitrary adiabatic splittings,
intra-well and inter-wells dynamics depends on the both adiabatic potentials
(i.e. on tunneling and adiabatic splitting).
Having in mind applications, the studies of these questions may concern the 
various molecular
systems undergoing so-called conversion of electronic states, 
isomerization reactions, or radiationless transitions arising from ''intersystem''
crossings of potential energy surfaces in molecular spectroscopy and chemical dynamics,
or inelastic atomic collisions. 

It is worth noting that there exists a huge literature devoted
to different approaches have been made by other authors 
to
the problem of crossing diabatic potentials
(see e.g. \cite{NU84}, \cite{BE84}, \cite{ZN94} - \cite{GS92}, 
\cite{PK61} - \cite{VG96})
but some important differences to
our work should be noted.
First, suffice it to say, that the problem how diabatic potentials crossing
modifies the adiabatic potentials (occurring as a result of this crossing)
quantization rules, has not been investigated at all. One of the
reason is that for many say standard rigid molecules with quite large
adiabatic splitting of energy levels, one may safely neglect any influence
of the upper adiabatic potential (i.e. to use the standard quantization rules).
However, nowadays the increasing precision of experimental tests of
molecular tunneling splitting and decay (and besides investigations
and synthesis of more and more new non-rigid molecules), makes the study of this
problem relevant and actual. 
Second, is a methodic note.
All previous approaches were based on the general semiclassical WKB formalism.
The crucial point to treat quantization for crossing diabatic potentials
is how to compute the contribution coming from
the contour around a complex turning point. The
accuracy of the WKB method can be improved considerably, 
\cite{PK61} - \cite{OV65} (more recent references on so-called Laplace contour integration can 
be found also in \cite{NU84}, \cite{MC03}) by the
appropriate choice of the integration path around 
the turning point. This method ascending to Landau \cite{LL65} 
appears to be quite accurate 
for the tunneling and over-barrier regions, (however,
even in this case there are some non-negligible 
corrections found in the papers \cite{ZN94}, \cite{ZN92},
\cite{ZN93}), but in the intermediate energy region
(where there are relevant contributions from all four quantum states occurring 
at the crossing diabatic potentials) the
method
becomes completely non-adequate. 
Besides the choice of these additional special trajectories (which one has to include to improve the
accuracy of the WKB method, and along which the semiclassical
motion is described by the Weber functions) depends on the detail form of the potential far from the
top, and therefore for each particular case the non-universal procedure should be perform from the very
beginning.

The essential simplification of the procedure can be achieved 
using                                              
the standard WKB semiclassical
approximation in the momentum space representation was also proposed
in the literature \cite{NU84}, \cite{ZN94}, \cite{BN65} - \cite{ZN93}. 
The method works well to compute the LZ transition probability, however, the application
of this approach to the level quantization problem
is difficult to realize. Indeed the problem requires to know 
the eigenfunctions in the coordinate space, and one can not bluntly use the Fourier
transform of the functions found in the momentum space, since the WKB method
gives us only asymptotics of the eigenfunctions.
These say drawbacks of the WKB-like methods did not allow to study
level quantization for crossing diabatic potentials in the previous
investigations, and, besides,
we believe we are the first to explicitely addresses the question on the behavior in
the intermediate
energy region. 
In all previous publications this region was considered as a very narrow and insignificant
one, or in the best case the results were obtained by a simple interpolation from the tunneling 
(with monotonic decay of
the transition probability) 
to the over-barrier (with oscillating behavior) regions.

Recently we have shown \cite{BV02} that 
semiclassical solutions of many eigenvalue problems
can be considerably simplified by including into the consideration
second order turning points. The fact is that one second order
turning point replaces two close linear turning points. Moreover
it turns out (see below) that connection matrices which link on the complex plane 
the solutions 
to the Schr\"odinger equation in the vicinity of the the crossing point
with the asymptotic solutions far from this point, can be calculated
from the solutions of the Weber equation. Increasing and decreasing
solutions in the classically forbidden region around a second order
turning point, are characterized by the action which has a minimum along
a certain trajectory, we will refer in what follows as the instanton
trajectory. The same kind of the Weber equation can be formulated
to calculate the connection matrices in the vicinity of a saddle point
(or a maximum) of the potential, but besides we should also relate
increasing and decreasing solutions at the crossing point (see also
our recent publications on LZ crossing phenomena \cite{BV03}).

Our aim is to construct semiclassical wave functions. To do it we use
connection matrix methodology which can be applied to any semiclassical approximation, 
but details of
the method depends on the order of the turning points. For the second order turning points which are
minima of a potential, the whole procedure is equivalent to the traditional instanton approach, 
and the imaginary time (i.e. after Wick rotation) instanton trajectories correspond to the
periodic orbits between the turning points, and the connection matrices in this case 
(see below and appendix \ref{A} to the paper) are real-valued ones.
It is not the case for the second order turning point which is the potential maximum. 
The complex-valued
connection matrix links two regions of infinite motion. Formally one might
refer 
the corresponding
wave
functions also as instanton ones applying twice the Wick rotations. LZ
crossing points are combinations of two second order turning points with 
two different Stokes
constants \cite{HE62} corresponding to one minimum and one maximum of 
the potential. In the tunneling
region there
exist periodic orbits for two solutions while two others correspond to unrestricted 
(infinite) motions. As above we will call these wave functions as instanton ones 
since they are Weber functions (like in
the traditional
instanton method) but with complex-valued arguments occurring as a result of complex coordinate frame
rotation.

Thus our approach in this paper 
is based on the
minimization of the functional of classical action 
in the upside-down potential, so-called instanton type approach,
which represents the most important area of the configurational
space where the semiclassical wave functions are localized ( 
see \cite{BM94}, \cite{MC03} - \cite{BV03}). 
The whole analysis can be brought into a more elegant form by introducing connection
matrices which link on the complex plane the semiclassical solutions 
to the Schr\"odinger equation for the
model potential of the problem under study and the exact solutions of the
so-called comparison equation which is valid near the crossing point,
where one can approximate the potential by linear or parabolic ones. The
explicit calculations of the connection matrices are rather involved since 
one should treat the four fundamental solutions to the left and to the right 
regions with respect to turning or crossing
points. Therefore the connection matrices, we are looking for, are $4 \times 4$ matrices. 
Although the
generalization for our case of the known already $2 \times 2$ connection matrices (see e.g., \cite{HE62}) 
is straightforward, it deserves some
precaution as it implies quite different procedures for the energy  (more accurately for $E/\gamma $
were $\gamma \gg 1$ is the
semiclassical parameter, see below) 
smaller (the tunneling region), larger (the over-barrier region), or of the order (the intermediate
region) of the potential barrier energy.
Within the framework of the connection matrix 
approach we present a full and unified description of
$1D$ (which is very often can be quite reasonable
approximation for real systems)
level quantization problem for diabatic potentials crossing. 

The remainder of our paper has the following structure. Section \ref{bas} contains basic methodical
details and equations necessary for our investigation. 
Except for a mathematical trick that eliminates a large amount
of tedious algebra and allows us to construct a regular
method for
calculating higher order perturbative corrections, the section
contains already known results.  
New physical results are collected in sections \ref{res}, \ref{inter} and
\ref{bath}, and partially in section \ref{mat}, where we calculate 
all needed connection matrices, which provide a very efficient method of finding 
semiclassical solutions to
the Schrodinger equation in potentials having several turning points. 
The knowledge of the connection matrices is important and significant 
not only in itself but also for developing a good analytical
approximation, and standardized numerical procedures.                                                               
In Section \ref{res} we find the 
quantization rules for the tunneling and over-barrier energy regions.
Section \ref{inter} is devoted to the intermediate
energy region where all four states occurring at the diabatic
potentials crossing should be regarded on the same footing. 
Different particular cases, depending on
the ratio of the model parameters are also examined in this section. 
In section \ref{bath} we investigate the linear coupling of the LZ
system to harmonic phonons and find that
it renormalizes of the parameters entering
the initial diabatic potentials crossing problems considered in the previous
sections \ref{res} - \ref{inter}.
The
last section \ref{discus} contains summary and discussion. 
In two appendices to the paper we collect some more specialized
technical material required for the calculations connection matrices
in different energy windows (appendix A), and to reduce fourth order Schr\"odinger equation
to two independent second order Weber equations (appendix B). Those readers
who are not very interested in mathematical derivations can skip these appendices finding
all essential physical results in the main text of the paper.

\section{Fourth order comparison equation for the crossing point.}
\label{bas}     

To move further on smoothly let us describe first our strategy.
First we should define all notations and relevant
points  of the diabatic potentials crossing problem.
We depicted the typical situation in the vicinity of the diabatic potentials
crossing point in Fig. 1. 
The diabatic potentials (1,2) are shown by thin solid lines, the adiabatic potentials
(3,4) by bold solid lines. Besides we have introduced in the picture
the adiabatic coupling energy $U_{12}$, the crossing point energy $U^{\# }$, and
$E_0$ is the characteristic zero-point oscillations energy
in the parabolic barrier approximated the lower adiabatic potential
near its top.

As a model for diabatic potentials in this paper we choose two parabola
$U_L$, $U_R$ with a symmetrical crossing in the point $x=0$.  
To be specific: let us consider two types of the diabatic 
potential crossing depicted in the Fig. 2. The corresponding adiabatic potentials
are, respectively, the double well or decay lower potential, and the
one-well upper adiabatic potential.
At arbitrary values of the parameter $U_{12}$ to find eigenstates and eigenfunctions for our
model potential we should solve the coupled Schr\"{o}dinger equations 
\begin{eqnarray} 
\label{nn3}
-\frac{1}{2}\frac{d^2 \Theta _L}{d x^2} + \gamma ^2 (U_L(x) - E) \Theta _L = \gamma ^2 U_{12}
\Theta _R \, ; \,
-\frac{1}{2}\frac{d^2 \Theta _R}{d x^2} + \gamma ^2 (U_R(x) - 
E ) \Theta _R = \gamma ^2 U_{12} \Theta _L \,
.
\end{eqnarray} 
Here $\gamma \gg 1$ is the
semiclassical parameter 
which is determined by the ratio of the characteristic potential scale over the
zero oscillation energy (i.e. $\gamma \equiv m \Omega _0 a_0^2/\hbar $, where $m$ is a mass of a
particle, $a_0$ is a characteristic length of the problem, e.g. the tunneling distance, $\Omega _0$ is a
characteristic frequency, e.g. the oscillation frequency around the potential minimum).

These equations (\ref{nn3}) can be written as one fourth order equation, e.g. for $\Theta _L$ 
\begin{eqnarray} 
\label{nn4} \frac{d^4
\Theta _L}{d x^4} - 2 \gamma ^2 (U_L(x) + U_R(x) -2E)
\frac{d^2 \Theta _L}{d x^2} - 4 \gamma ^2
\frac{d
U_L}{d x}\frac{d \Theta _L}{d x} + 4\gamma ^4 \left [(U_L - E)(U_R - E) - U_{12}^2 -
\frac{1}{2\gamma
^2}\frac{d^2 U_L}{d x^2}\right ] \Theta _L = 0  
\end{eqnarray} 
In what follows 
we use  $\Omega _0$ and $a_0$ to set corresponding 
dimensionless       
scales, e.g dimensionless energy $\epsilon = E/\gamma \Omega _0$,
$u_{L/R} = U_{L/R}/(\gamma \hbar \Omega _0)$, $u_{12} = 2 U_{12}/(\gamma \hbar \Omega _0)
$ (we introduce factor 2 in $u_{12}$ for ease of writing following below
equations), coordinate $X = x/a_0$, and we  
put $\hbar = 1$ 
(except where explicitely stated to the contrary
and the occurrences of $\hbar $ are necessary for understanding).

Luckily the equation (\ref{nn4}) admits semiclassical solutions by 
Fedoryuk method \cite{FE64} -
\cite{FE66} since the coefficients at the $n$-th order derivatives proportional to 
$\gamma ^{-n}$, and
therefore so small that higher order derivatives of the prefactor
(in the semiclassical form the wave function can be always presented as the prefactor
times the exponent) can be safely neglected
in finding of asymptotic solutions.  
Fedoryuk method makes possible to find asymptotic solutions
to the ordinary differential equations of the following form
\begin{eqnarray} 
\label{f1} 
y^{(n)} + \sum _{k=0}^{n-1} \gamma ^{n-k} f_k(X) y^{(k)} = 0
\, ,
\end{eqnarray} 
where we designated $y^{(k)} \equiv d^k y/dX^k$, and the coefficients
at the derivatives $f_k(X)$ are arbitrary functions of $X$.
Note that Eq. (\ref{nn4}) for $\Theta $ has this Fedoryuk form.
By the standard semiclassical substitution $ y = A \exp (\gamma W(X))$
(\ref{f1}) can be reduced to the set of equations combining
the terms proportional to $\gamma ^n$, $\gamma ^{n-1}\, \cdots $,
which for $\gamma \gg 1$ can be represented in the form 
of generalized so-called Hamilton - Jacobi and transport equations,
respectively
\begin{eqnarray} 
\label{f2} 
F(\lambda ) = \lambda ^n + \sum _{k=0}^{n-k} f_k(X)\lambda ^k  = 0 
\, ,
\end{eqnarray} 
and
\begin{eqnarray} 
\label{f3} 
\frac{d F}{d \lambda }\frac{d A}{d X} + \frac{1}{2}\frac{d^2 F}{d \lambda ^2}
\frac{d \lambda }{d X} A  = 0 \, ,
\end{eqnarray} 
where $\lambda = -\gamma d W/ d X$.

Noting 
that in the vicinity of the crossing point $X=0$ the parabolic diabatic potentials
can be replaced by the linear ones counted from the barrier top $U^\# $ 
\begin{eqnarray}
\label{nn6} u_{L/R}(X)  = u^\# \pm f X \, , 
\end{eqnarray} 
(as above $u^\# =U^\#/(\gamma \hbar \Omega _0)$), and
eventually the equation (\ref{nn4}) can be
presented into a more compact and simple form 
\begin{eqnarray} 
\label{nn7} 
\frac{d^4 \Theta _L}{d X^4} -
2\gamma ^2 \alpha \frac{d^2 \Theta _L}{d X^2}  
-2\gamma ^2 f \frac{d \Theta _L}{d X} +
\gamma ^4 [\alpha ^2 - f^2 X^2 - u_{12}^2] \Theta _L = 0 
\, ,
\end{eqnarray} 
where in our dimensionless units
$\alpha = 2(u^\# - \epsilon ) $.

The roots of the characteristic polynomial for (\ref{nn7}) 
\begin{eqnarray} 
\label{nn8} 
F(\lambda , X) =
\lambda ^4 - 2\gamma ^2 (u_L + u_R - 2 \epsilon )\lambda ^2 - 
4\gamma ^2 \frac{du_L}{d X}\lambda + 4\gamma
^4 \left
[(u_L - \epsilon )(u_R - \epsilon ) - u_{12}^2 - 
\frac{1}{2\gamma ^2}\frac{d^2 u_L}{d X^2}\right ] = 0 \, 
,
\end{eqnarray}
or in the equivalent form
\begin{eqnarray} 
\label{b2} 
F(\lambda ) =
\lambda ^4 - 2 \alpha \gamma ^2\lambda ^2 - 2 \gamma ^2 f \lambda +\gamma ^4(\alpha ^2 - 
u_{12} ^2 - f^2
X^2) \,
, 
\end{eqnarray} 
determine independent
solutions to (\ref{nn7}).
Solving the equation $F(\lambda ) = 0$ perturbatively over $\gamma ^{-1} \ll 1$ we find 
\begin{eqnarray}
\label{b3} 
\lambda _j = \lambda _j^0 + \lambda _j^1 \, , 
\end{eqnarray} 
where 
\begin{eqnarray} 
\label{b4} 
\lambda
_j^0 = \pm \gamma 
\left [
\alpha \pm \sqrt {u_{12} ^2 + f^2 X^2}\right ]^{1/2} \, , 
\end{eqnarray} and
\begin{eqnarray} 
\label{b5} 
\lambda _j^1 = \pm \frac{f}{2 \sqrt {u_{12} ^2 + f^2 X^2}}
\,  
\end{eqnarray} 
we find finally the four asymptotic solutions of (\ref{nn7}) 
\begin{eqnarray}
\label{b6} 
\{ y_j \} \equiv \{ \Theta _+^+ , \Theta _+^- , \Theta _-^+ , \Theta _-^- \} = 
\left (\frac{d F}{d \lambda }\right )^{-1/2} \exp \left [\int _{0}^{X} \lambda _j(X^\prime ) 
d X^\prime \right ] \, . 
\end{eqnarray} 
The
subscripts in (\ref{b6}) corresponds to the upper or lower adiabatic levels, and the superscripts
are referred to the sign of the action.

As it was mentioned above in the vicinity of the crossing point one can replace (\ref{nn4}) by
(\ref{nn7}), and by the substitution 
\begin{eqnarray} 
\label{nn11}
\Theta _L = \exp (\kappa _{1 , 2} X) \Phi _L^{1 , 2} \, , 
\end{eqnarray} 
we can find
the equation for $\kappa $  
\begin{eqnarray} 
\label{b8} 
\kappa ^4 - \alpha \gamma ^2\kappa ^2 + \frac{1}{4} \gamma ^4 u_{12}^2 = - \kappa ^4 \delta ^2(1 + 2\delta )
+ R(\kappa , \delta )\, ,
\end{eqnarray} 
where
\begin{eqnarray}
\nonumber
R(\kappa , \delta ) = (2 \kappa ^6)^{-1}(1-3\delta )(1+\delta )^{-3}(1 - Q - \sqrt {1 - 2Q^2)})\, ; \, 
Q=8\delta ^2(1+\delta )\, ,
\end{eqnarray}
\begin{eqnarray} 
\label{b9} 
\delta = \frac{\gamma ^2 f}{4\kappa ^3 } < \frac{1}{4} \, , 
\end{eqnarray} 
and
\begin{eqnarray} 
\label{b10} 
\kappa _{1 , 2} = \pm \gamma \sqrt \alpha \left (1 - \frac{\delta ^2}{2}\right )  \, . 
\end{eqnarray} 
It can be proved that the 4-th order equation (\ref{nn7}) in variables (\ref{nn11}) is reduced to the equation
with constant coefficients in front of all derivatives and with the free term in the form
of a quadratic over $X$ function. In the case when the exponent in (\ref{nn11}) is a solution
to the equation (\ref{b8})the transformed equation is reduced to the Weber equations
upon neglecting anharmonic terms like $X^2 dF/dX$, $X^3 F(X)$, and $X^4 F(X)$.
We presented all details of this reduction in appendix \ref{B} to the paper.
Thus the equation (\ref{nn7}) is reduced to two independent Weber equations with the 
known
fundamental solutions \cite{EM53} 
\begin{eqnarray} 
\label{nn13} 
\{\Theta _L\} = \left \{ \exp( \pm \gamma
\sqrt \alpha X) D_{-\nu } \left (\left (\frac{f^2\gamma ^2}{\alpha }\right )^{1/4}X\right ) \, , \, 
\exp(
\pm
\gamma \sqrt \alpha X) D_{-1 - \nu } \left (\left (\frac{f^2\gamma ^2}{\alpha }\right )^{1/4}X\right
) \right
\} \, ,
\end{eqnarray} 
where $\nu = \gamma u_{12}^2/(4 f \sqrt \alpha )$
is so-called Massey parameter. The corrections to the indices of the parabolic cylinder
functions $D$ and to the
arguments of these functions can be found from (\ref{b8}).

Presented above the leading terms of these solutions 
corresponding to the tunneling case, i.e. 
(here we use
a dimensional energy
$E$)
\begin{eqnarray} 
\label{j1} 
E < (U^\# - U_{12}) 
\, 
\end{eqnarray} 
(in our dimensionless units it is $\alpha > u_{12}$), 
where the characteristic fourth order polynomial (\ref{nn8}) 
can be
reduced to the second order one (i.e. two pairs of roots are nearly degenerated),
are known in the literature
(see e.g. \cite{LL65} - \cite{BM94}) but the Fedoryuk method we used, 
gives us also in the tunneling
region the
higher
order over the parameter $\delta $ (\ref{b9}) corrections.
 
In the tunneling region (\ref{j1}) one can expand the roots of (\ref{b8}) in
terms
of the parameter $\delta $ (\ref{b9}). 
Using the substitution (\ref{nn11}) 
to transform (\ref{nn4}) 
we can find easily that at the conditions (\ref{b8}), (\ref{b9}) the
coefficients at the fourth and at the third order derivatives in the transformed
fourth order differential equation for $\Phi $ are small
(proportional to
$\delta $ and to $\sqrt \delta $ respectively) and thus this fourth order equation can be
rewritten
as two second order Weber equations with the solutions 
$$ 
D_{p^{(1 , 2)}}(\beta X) \, , 
$$
where 
\begin{eqnarray} 
\label{b11} 
p^{(1)} = - 1 + \frac{\delta }{4} - \nu \, , \, p^{(2)} = \frac{\delta }{4} - \nu \, , 
\beta  = \left (\frac{\gamma ^2 f^2}{\alpha }\right )^{1/4} \left (1 +\frac{\delta ^2 }{4} \right ) \, .
\end{eqnarray} 

The same manner in the over-barrier energy region 
i.e. 
\begin{eqnarray} 
\label{j2} 
E > (U^\# + U_{12}) \, 
\end{eqnarray} 
(again as in (\ref{j1}) we have used dimensional
units, and in dimensionless form (\ref{j2}) reads as $-\alpha > u_{12}$),
when the energy is larger than the upper adiabatic potential
minimum,
the roots of the equation
(\ref{nn8}) are complex - conjugated and having the same structure as presented
above (see also (\ref{apen6}) in the appendix \ref{B}) 
for the tunneling region
with the roots $\kappa $ given 
\begin{eqnarray} 
\label{b12} 
\kappa _{1 , 2} = \pm i \gamma \sqrt {|\alpha |}\left (1 - \frac{\tilde {\delta }^2}{2}\right ) \, .
\end{eqnarray}
Besides in (\ref{apen6})
\begin{eqnarray}
\nonumber
\kappa _0  = i \frac{\gamma }{\sqrt 2}(|\alpha | + \sqrt {\alpha ^2 - u_{12}^2})  
\, ,
\end{eqnarray}
and with $\tilde \delta $ playing the role of the small parameter in this region 
\begin{eqnarray}
\label{b13} 
\tilde \delta = \frac{f}{4 \gamma |\alpha |^{3/2}} \, . 
\end{eqnarray} 
Again as above for the
tunneling region, the coefficients at the higher order derivatives are small, and therefore, the function
$\Phi $ (\ref{nn11}) satisfies the Weber equation with the fundamental solutions 
$$ 
D_{\tilde p^{(1 ,
2)}}(\tilde {\beta }_{(1 , 2)} X)  \, , 
$$ where 
\begin{eqnarray} 
\label{b14} 
\tilde p^{(1)} = - 1 - i\frac{\tilde \delta }{4} + i \tilde \nu \, , \, 
\tilde p^{(2)} =
i\frac{\tilde \delta }{4} + i \tilde \nu \, , \, 
\\ 
\nonumber
\tilde {\beta }_1 = \exp(i\pi /4) \left (\frac{\gamma ^2 f^2}{|\alpha |}\right )^{1/4}\left (1 + \frac{\tilde {\delta }^2}{4}\right ) \, , \, 
\tilde {\beta }_2 = \exp(-i 3\pi /4) \left (\frac{\gamma ^2 f^2}{|\alpha |}\right )^{1/4}\left (1 + \frac{\tilde {\delta }^2}{4}\right ) 
\, 
\end{eqnarray}
($\tilde \nu $ is defined as the Massey parameter entering (\ref{nn13}) with $\alpha \to |\alpha |$, i.e.
$\tilde \nu = (\gamma u_{12}^2/(4 f \sqrt {|\alpha |})$). 
Like it was
for the tunneling region (\ref{b11}), the leading terms of the expansion (\ref{b14}) coincide with the
well - known results, but from (\ref{b14}) we are able to compute the
corrections to the
main terms.

The analogous task for the
intermediate energy region, i.e. (in dimensional units)  
\begin{eqnarray} 
\label{j3} 
(U^\# + U_{12}) \geq E \geq (U^\# - U_{12}) \, ,
\end{eqnarray} 
is much more tricky.
Our results will be presented in section \ref{inter}, but a few comments are
necessary here.
In the problem we have three dimensionless parameters 
characterizing the energy ($\alpha $),
the level coupling ($u_{12}$), and the potential ($f$), and besides for the ease of semiclassical
estimations we keep also the semiclassical parameter $\gamma \gg 1$. Note also that these parameters
are not independent ones, and the relation $u_{12}=2f^2$ (which we will be useful
in our further consideration) should be satisfied. 
In terms of these parameters within the intermediate energy region (\ref{j3}),
we have the subregion, $S^\prime $, $|\alpha | \leq 2 \gamma ^{-1}$, and $u_{12} \leq 2 \gamma ^{-1}$,
and the intermediate subregion, $S^{\prime \prime }$, where $\gamma \sqrt {u_{12}}/2 \gg 1$. 
In the section \ref{inter} we calculate the connection
matrices for the both subregions, and details of the reduction procedures, which are different in
$S^\prime $ (where the comparison equations are reduced to two decoupled Airy equations)
and $S^{\prime \prime }$ (where these comparisons equations are Weber ones)
are described in the appendix \ref{B}.

\section{Connection matrices.}
\label{mat}     

The purpose of this section is to briefly indicate the main
steps in the derivation connection matrices.
The matching points we must find to quantize the energy
levels depend essentially on the energy window under consideration (\ref{j1}), (\ref{j2}), (\ref{j3}).
The tunneling region is placed in the lower adiabatic potential.
In the WKB method in this case starting from the crossing point ($X=0$) one has to
investigate the classically forbidden region in the lower adiabatic
potential barrier (see Fig. 3a and the corresponding figure caption for all notations). 
The solutions can
be found easily in the vicinity
to the crossing point but to derive the quantization
rules one should know also the solutions quite far from the crossing point.
To do it explicitely in the WKB method we should match the two exponentially
decreasing and two exponentially increasing solutions in the barrier with the oscillation
solutions in the wells. 
Technically the matching should be performed asymptotically, i.e. at small
$|X|$ but for large enough $\sqrt \gamma |X|$. To do it one 
has to  
calculate all needed connection matrices (namely at the crossing point, and at
the linear and second order turning points,  and the shift
matrices from the crossing point to the turning points in the classically
forbidden region and between the turning points in the classically
accessible region). Within the instanton type method the trajectory goes 
through only the classically
forbidden region
(see Fig. 3b),
and matching should be performed only at two second order turning points.
In the over-barrier energy region (see Fig. 4), the matching 
is performed by using the crossing point connection matrix $\hat {U}^\prime _c$, the shift matrix
$\hat L$ connecting the crossing
point and the linear turning points at the upper adiabatic
potential, and the shift matrices $\hat {L}_{L/R}$ belonging to the diabatic potentials.
In this case all matching solutions are oscillating ones. 
Finally for the intermediate
energy region no real-valued turning points for the upper states (see Fig. 5) 
and the matching between two oscillating and two exponentially varying
solutions
is determined by the connection matrix (see Fig. 5) linking the linear 
imaginary turning
points of the adiabatic potentials.

To recast the analysis into a compact
form it is convenient to formulate the general procedure for calculating of 
the connection matrices for an
arbitrary combinations of the first and of the second order turning points. 
After that the procedure can
be applied to any particular problem under investigation. To do it, one has to extend the
known for linear turning points procedure \cite{HE62}. 
For a generic semiclassical equation 
\begin{eqnarray} 
\label{p1} 
\frac{d^2 \Psi }{d z^2} + \gamma ^2 q(z) \Psi (z) = 0 \, ,
\end{eqnarray} 
in the limit $\gamma \gg 1$ the Stokes and anti-Stokes lines are determined
by the following conditions, respectively 
\begin{eqnarray} 
\label{pp1} \Re W(z) = 0 \, , 
\end{eqnarray} 
and
\begin{eqnarray} 
\label{pp2} 
\Im W(z) = 0 \, , 
\end{eqnarray} 
where the action
\begin{eqnarray} 
\label{p2} 
W(z) = \int ^{z} \sqrt {q(z)} dz \, . 
\end{eqnarray} 
The lower integration limit in (\ref{p2}) is not relevant
because we are interested in semiclassical solutions for
large $|z|$.
These Stokes and anti-Stokes lines separate the complex plane $z$
into the sectors. On the anti-Stokes lines the increasing
and decreasing solutions become equal, and the type of the solutions
is interchanged upon crossing of the anti-Stokes lines. The Stokes
lines are bisectors between neighboring anti-Stokes lines. After
the crossing with the Stokes line, one should add to the coefficient
at the decreasing solution the coefficient at the increasing solution
times so-called Stokes constant. The latter one is occurred as a result
of going around the turning point and depends on the turning point 
type.

To find the connection matrices 
for the tunneling region we have to
establish the correspondence between the solutions of the 
fourth order differential equation (\ref{nn4})
and those for the localized in the left ($L$) 
and in the right ($R$) wells states. 
In the case $\alpha \gg
f|X|$ for the diabatic potentials, 
the action can be computed starting from the both wells ($R$ and $L$)
\begin{eqnarray} 
\label{b15} 
\gamma W_L^* \simeq \gamma W_{0 L}^* + k_0 X + \frac{\beta ^2}{4} X^2 \, , \gamma
W_R^* \simeq \gamma W_{0 R}^* - k_0 X + \frac{\beta ^2}{4} X^2 \, , 
\end{eqnarray} where $k_0 = \gamma \sqrt \alpha $ is imaginary momentum (
$U^{\# }$ sets the energy corresponding to the diabatic potentials
crossing point), 
and $W_0^{L , R}$ are the actions
computed from an arbitrary distant point in the $L$ 
or in the $R$ wells, respectively to the point $X=0$.
From the other hand in the adiabatic potentials 
$u^\pm = u^\# \pm \sqrt {u_{12}^2 + f^2 X^2}$ the
corresponding actions can be represented 
\begin{eqnarray} 
\label{b16} 
\gamma W^\pm - \gamma W_0^\pm = k_0X
\pm \frac{\beta ^2}{4} X^2 sign (X) \, . 
\end{eqnarray} 
Explicitely comparing the semiclassical wave
functions in the both representations 
(adiabatic and diabatic ones) it is easy to see that the adiabatic
functions in the potential $U^-$ coincide 
with the diabatic functions for localized $L$ and $R$ states at
$X<0$ and $X>0$ respectively. 
The adiabatic functions for the upper potential $U^+$ correspond to the
tails of the diabatic wave functions localized in the opposite wells.  
Therefore in the level crossing
region the $L/R$ diabatic functions are transformed into the $R/L$ functions, 
and the interaction
entangles the diabatic states with the same sign of $k_0 X$. 
Thus we have only four non-zero amplitudes of
the following transitions 
\begin{eqnarray} 
\label{b17} 
\langle \Phi ^+_L|\Phi ^-_R\rangle \, , \, \langle
\Phi ^-_L|\Phi ^+_R\rangle \, , \, \langle \Phi ^+_R|\Phi ^-_L\rangle \, , \, \langle \Phi ^-_R|\Phi
^+_L\rangle \, . 
\end{eqnarray} 
Recalling that 
\begin{eqnarray} 
\label{b18} 
\gamma W^\pm = \gamma \int
\left (\alpha \pm \sqrt {u_{12} ^2 + f^2 X^2}\right )^{1/2}
\simeq k_0
X \pm \frac{\beta ^2}{4} X^2 \pm \frac{\nu }{2}(1 - \ln \nu ) \, , 
\end{eqnarray} 
we come to the
conclusion that the quantum solutions (\ref{b11}), valid in the vicinity of the level crossing point
asymptotically, match smoothly increasing and decreasing solutions, and it leads to the Landau
description \cite{LL65} of the level crossing transitions.
To illustrate the presented above analytical results we show schematically in Fig. 6
the matching of the asymptotic (Fedoryuk) solutions (\ref{b6}) for the crossing 
diabatic potentials with localized initial and final states via the Weber functions in the
tunneling region. We use the symmetric basis constructed from the
functions (\ref{b6}) (see detail description in the Fig. 6 caption). 

In the tunneling region (\ref{j1}) for
every well ($L$ or $R$) there exist increasing and decaying exponentially 
real-valued solutions to the
Schr\"odinger equation. The solutions are matched at the crossing point, 
therefore they are linked by the
real-valued $4 \times 4$ connection matrix which should have two 
$2 \times 2$ blocks linking the
increasing (decreasing) diabatic solution in the $L$-well with the 
decreasing (increasing) diabatic
solution in the $R$-well, in the agreement with the standard 
Landau scheme of the tunneling transitions
\cite{LL65}. Omitting a large amount of tedious algebra we can represent 
the connection matrix linking the
''asymptotic'' (i.e. in the left/right ($L$, $R$) wells and for 
the upper/lower ($+$, $-$) adiabatic
potentials) solutions in the tunneling energy region in the following form 
\begin{eqnarray} 
&& 
\label{q1}
\left ( 
\begin{array}{c} 
\Phi _R^- \\ \Phi _R^+ \\ \Phi _L^+ \\ \Phi _L^- 
\end{array} 
\right ) 
= \left (
\begin{array}{cc} 
\hat {M}_c^{(+)}\hat {L}^{(c)}_R\hat {M}_c^{(-)}\hat F_c & 0 \\ 0 & \hat 1 
\end{array}
\right ) 
\hat {U}_c 
\left ( 
\begin{array}{cc} 
\hat F_c\hat {M}_c^{(+)}\hat {L}^{(c)}_L\hat {M}_c^{(-)} & 0
\\ 0 & \hat 1 
\end{array} 
\right ) \left ( 
\begin{array}{c} 
\Phi _L^+ \\ \Phi _L^- \\ \Phi _R^- \\ \Phi
_R^+ 
\end{array} 
\right ) \, . 
\end{eqnarray} 
Here $\hat U_c$ is the $4 \times 4$ connection matrix at the
crossing point, which in the tunneling region has the following form 
\begin{eqnarray} && 
\label{q8} 
\hat
{U}_c = \left [ 
\begin{array}{cccc} p & 0 & 0 & -\cos (\pi \nu ) \\ 0 & (\sin^2 (\pi \nu ))/p & - \cos
(\pi \nu ) & 0 \\ 0 & \cos (\pi \nu ) & p & 0 
\\ \cos (\pi \nu ) & 0 & 0 & (\sin ^2(\pi \nu ))/p
\end{array} 
\right ] \, , 
\end{eqnarray} 
where we designated 
\begin{eqnarray} 
\label{xx} 
p=\frac{\sqrt
{2\pi }\exp(-2 \chi )}{\Gamma (\nu )} \, , 
\end{eqnarray} and 
$ \chi = (\nu /2) - (1/2)\left (\nu - (1/2)
\right ) \ln \nu $.  
The matrices $\hat {M}_c^{(+)}$ and $\hat {M}_c^{(-)}$ are the $ 2 \times 2$
connection matrices at the corresponding turning points, which are determined 
by the phase shifts at these
points 
\begin{eqnarray} && 
\label{q2} 
\hat {M}_c^{(-)} = \left ( 
\begin{array}{cc} 1 & -i \\ -(i/2) &
(1/2) \end{array} 
\right ) \, , 
\end{eqnarray} 
and $\hat {M}_c^{(+)}$ is the matrix Hermitian conjugated
to (\ref{q2}). 
The $\hat L^{(c)}_{L/R}$ and $\hat F_c$ matrices are called shift 
matrices, and those are
related to the variations of the coefficients of increasing and 
decaying semiclassical solutions in the
regions between the turning points ($\hat F_c$ is the shift matrix 
when one moves from the crossing to the
turning point in classically forbidden region, 
and $\hat L^{(c)}_{L/R}$ are the shift matrices in the
classically accessible regions). Explicitely we get 
\begin{eqnarray} 
&& 
\label{q3} 
\hat F_c = \left(
\begin{array}{cc} 
\exp (-\gamma W_B^*/2) & 0 \\ 0 & \exp (\gamma W_B^*/2) \end{array} \right ) \, .
\end{eqnarray} 
Here $W^*_B$ is the action in the lower
adiabatic potential barrier. Finally the structure of the shift matrices 
$\hat {L}^{(c)}_{L/R}$ is
\begin{eqnarray} 
&& 
\label{q33} 
\hat {L}^{(c)}_{L/R} = \left( 
\begin{array}{cc} 
\exp (i \gamma W_{L/R}^*)
& 0 \\ 0 & \exp (-i \gamma W_{L/R}^*) 
\end{array} 
\right ) \, , 
\end{eqnarray} 
where $W_{L/R}^*$ is the
action calculated by the integration between the turning points. We present 
explicit expressions for the total connection matrix in
the appendix \ref{A}.
 
The same manner can be treated the over-barrier region (\ref{j2}). 
In this case the crossing
point is in the classically accessible region for the both potentials. 
The fundamental diabatic solutions
can be represented as the waves propagating in the opposite directions, 
and the complex-valued connection
matrix has as it was for the tunneling region $2 \times 2$ block structure, where the blocks link the
waves in the $L$ and in the $R$ wells propagating in the same direction. Specifically the corresponding
connection matrix at the crossing point $\hat {U}_c^\prime $ 
\begin{eqnarray} 
&& \label{q88} 
\hat
{U}_c^\prime = \left [ 
\begin{array}{cccc} 
s\exp (-i\phi ) & 0 & 0 & -\exp (-\pi \nu ) \\ 0 & s \exp (i
\phi ) & - \exp (-\pi \nu ) & 0 \\ 
0 & \exp (-\pi \nu ) & s \exp (-i \phi ) & 0 \\ 
\exp (-\pi \nu ) & 0 &
0 & s\exp (i \phi ) 
\end{array} 
\right ] \, , 
\end{eqnarray} 
where we denoted $ s = \sqrt {1 - \exp
(-2\pi \nu )}$, $\phi = \arg \Gamma (-i \nu ) + \Im (2 \tilde \chi )$, 
and 
\begin{eqnarray}
\label{cd}
\tilde \chi = - (i/2)((\pi
/4) + \nu (1 - \ln \nu)) + (1/4)(\pi \nu + \ln \nu )
\, ,
\end{eqnarray} 
should be multiplied by two blocks: the block from
the left gives the contribution at the turning point 
and includes the shift matrix to the crossing point
in $L$ and in $R$ wells of the lower adiabatic potential; the 
right block is related to the turning point
and to the shift matrix to the crossing point in the upper one-well 
adiabatic potential. Thus finally in
the over-barrier region we get 
\begin{eqnarray} 
&& 
\label{q9} \left ( 
\begin{array}{c} 
\Phi _R^- \\ \Phi
_R^+ \\ \Phi _L^+ \\ \Phi _L^- 
\end{array} \right ) = 
\left ( 
\begin{array}{cc} 
\hat {M}_c^{(+)}\hat
{L}^{(c)}_R & 0 \\ 0 & \hat {M}^{(+)}\hat L 
\end{array} 
\right ) 
\hat {U}_{c}^\prime \left (
\begin{array}{cc} 
\hat {L}^{(c)}_L\hat {M}_c^{(-)} & 0 \\ 0 & \hat {L}\hat {M}^{(-)} 
\end{array} 
\right ) 
\left (
\begin{array}{c} 
\Phi _L^+ \\ \Phi _L^- \\ \Phi _R^- \\ \Phi _R^+ 
\end{array} 
\right ) 
\, . 
\end{eqnarray}
Here we used the same notations as it was above for the tunneling region, and 
besides the matrices $\hat
M^{(\pm )}$ are transposed with respect to the matrices $\hat M_c^{(\pm )}$ 
given in (\ref{q2}), and the
new shift matrix $\hat L$ is 
\begin{eqnarray} 
&& 
\label{q333}
\left( 
\begin{array}{cc} 
\exp (-i \gamma W^*/2) & 0 \\ 0 & \exp (i \gamma W^*/2) 
\end{array} 
\right ) \,
, 
\end{eqnarray} 
(remind that $W^*$ is the action in the upper adiabatic potential). Combining altogether
(\ref{q2}) - (\ref{q333}) one can trivially find the 
full connection matrix
for the over-barrier energy region (\ref{j2}). 
We present the explicit form of the matrix in the appendix \ref{A}.
 
More tricky task is to calculate the connection matrix in the intermediate 
energy region (\ref{j3}). 
Following the same line as above we first present the general 
structure of the connection matrix in the
intermediate energy region 
\begin{eqnarray} && 
\label{q99} \left ( 
\begin{array}{c} \Phi _R^- \\ \Phi _R^+
\\ \Phi _L^+ \\ \Phi _L^- 
\end{array} 
\right ) = 
\left ( 
\begin{array}{cc} 
\hat {M}_c^{(+)}\hat
{L}^{(c)}_R\hat {M}_-^{(+)} & 0 \\ 0 & \hat {M}_+^{(-)} 
\end{array} 
\right ) 
\hat {U}_{c}^{\prime \prime } \left ( 
\begin{array}{cc}
\hat {M}_-^{(-)}\hat {L}_L^{(c)}\hat {M}_c^{(-)}
 & 0 \\ 0 & \hat {M}_+^{(+)}  
\end{array} 
\right ) \left ( 
\begin{array}{c} 
\Phi _L^+ \\ \Phi _L^- \\ \Phi _R^- \\ \Phi
_R^+ 
\end{array} 
\right ) \, . 
\end{eqnarray} 
These matrices $\hat {M}_\pm ^{(\pm )}$ have been introduced in our paper \cite{BV02}
for the imaginary turning points characterizing the both adiabatic potentials in the region
$|\alpha | < u_{12}$, and they read
\begin{eqnarray} && 
\nonumber  
\left ( 
\begin{array}{cc} 
1 & 0 \\ 
(i/2)\exp (- \gamma W_i^\pm ) & 0 
\end{array} 
\right ) 
\, , 
\end{eqnarray} 
where $W_i^\pm $ are so-called Euclidian actions in the reversed upper and lower
adiabatic potentials, which can be estimated as
$$
W_i^\pm \simeq \frac{\pi q_{1 , 2}}{\gamma }
\, ,
$$
where
$$
q_{1 , 2} = \frac{\gamma u_{12}\sqrt {u_{12} \pm \alpha }}{4 f} \, ,
$$
and all other matrices entering (\ref{q99}) are defined already.
Since $M_\pm ^{(\pm )}$ turn into the unit matrices at $\alpha > u_{12}$ and
$\alpha < u_{12}$, the connection matrix (\ref{q99}) matches continuously into the 
corresponding matrices (\ref{q1}), (\ref{q9}) for the tunneling and over-barrier  regions. 

The
connection matrix in the intermediate energy subregion $S^{\prime \prime } $
can be calculated using Weber function asymptotic expansion
for large complex indices (\ref{apen15}), which are the solutions to the comparison equations
(\ref{b8}) in the intermediate energy subregion $S^{\prime \prime } $.
These 4 roots are arranged clock-wise and counter clock-wise 
on the radius $\gamma \sqrt {u_{12}/2}$ circle around the crossing point.
The following combinations of the comparison equations match the semiclassical
solutions (\ref{b6})
$$
\Theta _1^+ + \Theta _4^+ \leftrightarrow \Theta _+^+ \, \; \,  
\Theta _2^- + \Theta _3^- \leftrightarrow \Theta _+^- \, \; \,
\Theta _1^- + \Theta _3^+ \leftrightarrow \Theta _-^+ \, \; \,  
\Theta _2^+ + \Theta _4^- \leftrightarrow \Theta _-^- \, .
$$ 
Combining together the asymptotic expansions for these
combinations, we find 
at the crossing point, the matrix 
$\hat U_c^{\prime \prime }$ is 
\begin{eqnarray} && 
\label{q888} 
\hat {U}_c^{\prime \prime } = \left [ 
\begin{array}{cc} 
(\sqrt {2\pi }/\Gamma (q^*))\exp (-2\chi (q^*)) & 0   \\
0 & (\Gamma (q)/\sqrt {2\pi })\exp (2\chi (q))(1 - \exp (-2\pi q_2)\cos ^2(\pi q_1)) \\
0 & \exp (-2\pi q_2)\cos (\pi q_1)   \\
\exp(-2\pi q_2)\cos (\pi q_1) & 0  
\end{array} 
\right .  
\end{eqnarray}
\begin{eqnarray}
\nonumber
\left .
\begin{array}{cc}
0 & \exp (-2\pi q_2)\cos (\pi q_1)  \\
-\exp (-2\pi q_2)\cos (\pi q_1)  &  
0  \\
(\sqrt {2\pi }/\Gamma (q))\exp (2\chi (q)) & 0 \\
0 & (\Gamma (q^*)/\sqrt {2 \pi })\exp (2\chi (q^*))(1 - \exp (-2\pi
q_2)\cos ^2(\pi q_1)) 
\end{array} 
\right ] \, , 
\end{eqnarray} 
where as above
\begin{eqnarray}
\label{a55}
q  = q_1 + i q_2\, ; \, 
q_{1 , 2} = \frac{\gamma u_{12}\sqrt {u_{12} \pm \alpha }}{4 f} \,
; \, q^* = q_1 - i q_2 \, ,
\end{eqnarray}
and, besides, we introduce the following abridged notations
\begin{eqnarray}
\label{a551}
\chi = \chi _1 + i \chi _2 \, ; \, 2 \chi _1 = q_1 - \left (q_1 - \frac{1}{2}\right ) \ln |q| + \varphi q_2 \, 
,
\end{eqnarray}
and analogously 
\begin{eqnarray}
\label{a552} 
2\chi _2 = q_2 - q_2\ln |q| - \varphi \left (q_1 - \frac{1}{2}\right ) 
\, , 
\end{eqnarray}
where $\varphi $ is defined by (\ref{apen10}). 
Now the full connection matrix in the both intermediate energy subregions 
can be found easily
simply collecting the given above expressions, and the explicit form
for the connection matrix is presented in the appendix \ref{A}.
Note that the intermediate energy region connection matrix (\ref{q888}) has the same
block structure as the connection matrices in the tunneling and in the over-barrier
regions. 
This is a consequence of the fact that in a neighborhood the diabatic potential crossing point,
only the Weber functions with equal indices can be hybridized. At $|\alpha | = u_{12}$ the
connection matrix (\ref{q888}) turns into the connection matrices (\ref{q8}) for the tunneling
region and into (\ref{q88}) for the over-barrier energy region,
and it enables us to construct semiclassical solutions for any arbitrary energy window.
Note, however, that in the intermediate energy region the Massey parameter is replaced by the complex index $q$.
In the section \ref{inter} we will present another derivation of the connection matrix (\ref{q888}),
and will discuss specific relations between the adiabatic and diabatic states in the intermediate
energy region.

\section{Quantization rules.}
\label{res}

In the tunneling energy region, one has only real-valued eigenfunctions, since
in the both wells there are only the localized states. 
In this energy window the connection matrix
linking the ''asymptotic'' (i.e. in the left/right ($L$, $R$) wells and for
the upper/lower ($+$, $-$) adiabatic potentials) solutions 
is represented in the form (\ref{q1}) and is given by Exps. (\ref{ap1}) - (\ref{ap1111}).
Within the WKB method we should match the two exponentially
decreasing and two exponentially increasing solutions in the barrier with the oscillation
solutions in the wells, thus it requires the knowledge
of the connection matrices at the crossing point and at
the linear turning points, and the shift
matrices from the crossing point to the turning points in the classically
forbidden region and as well between the turning points in the classically
accessible region (Fig. 3a). 
Within the instanton type method the trajectory goes through only the classically forbidden region
(see Fig. 3b),
and to perform the matching one should know  also the connection matrices
for the second order turning points.

The same manner can be treated the over-barrier region with the corresponding connection matrix 
(\ref{q9}), and (\ref{ap2}).
Evident modifications of the given above expressions for the tunneling and over-barrier
regions should be performed to treat the intermediate energy windows. Indeed
in this case one has to take into account also
the contributions from the imaginary turning points. The procedure is reduced to replacement
of oscillating factors by exponentially decaying ones (see details in the next section).
Finally for the intermediate
energy region no real-valued turning points for the upper states (see Fig. 5) 
and the matching between two oscillating and two exponentially varying
solutions
is determined by the connection matrix (\ref{q99})  
linking the linear turning
points of the adiabatic potentials (see Fig. 5).

\subsection{Two diabatic parabolic potentials crossing.}

Now 
(collecting the explicit expressions for all needed connection matrices
from the appendix \ref{A} to the paper)
we are in the position to derive the quantization rules,
which
can be formulated as a condition that the amplitudes of exponentially
increasing at $X > 0$, and $X< 0$, respectively, solutions $\Phi _L^+$, $\Phi _R^+$,
must be vanished. Taking into account that $W_L^* = W_R^*$ (the actions
in the corresponding wells of the lower adiabatic potential) and using
the connection matrix relating the fundamental solutions of the Weber equation, we
can formulate the corresponding quantization rule for the tunneling region
in terms of the matrix elements defined by (\ref{q1}) 
\begin{eqnarray}
\label{qq1} &&
m_{22}m_{33} - m_{23}m_{32} = 0
\, ,
\end{eqnarray}
where $m_{ij}$ are corresponding matrix elements from (\ref{q1}).

Putting all together we can find from (\ref{q1}) - (\ref{q8}), and (\ref{qq1})
the quantization rule for this case 
\begin{eqnarray}
\label{e1} &&
\tan (\gamma W_L^*)  = \pm \frac{2}{p}\exp (\gamma W_B^*)
\, ,
\end{eqnarray}
where $W_B^*$ is the action in the barrier formed in the lower adiabatic potential,
and $p \equiv U_{11}$ is the corresponding matrix element of the
connection matrix (\ref{q8}).

Only the factor $1/p$ varying from 0 to 1 in the diabatic and in
the adiabatic limits, respectively, makes
this quantization condition (\ref{e1}) different
from the well known \cite{LL65} quantization rule for
the symmetric double-well potential. Correspondingly, the tunneling
splitting at finite values of the so-called Massey parameter $\nu $ can be represented
as a product
\begin{eqnarray}
\label{e2} &&
\Delta _n = \Delta _n^0\, p(\nu )
\, ,
\end{eqnarray}
of the tunneling splitting $\Delta _n^0$ in the adiabatic
potential and the factor
\begin{eqnarray}
\label{e3} &&
p(\nu )= \frac{\sqrt {2\pi }}{\Gamma (\nu )} \gamma ^{\nu - (1/2)}\exp (-\nu )
\, ,
\end{eqnarray}
associated with the transition amplitudes between the diabatic
potentials in the crossing region.

It is particularly instructive to consider
(\ref{e1}) as the standard \cite{LL65} Bohr-Sommerfeld
quantization rule, where in the r.h.s. the both, geometrical
$\varphi _n$ and the tunneling $\chi _n$ phases are included additively.
In the adiabatic limit when $p(\nu ) \to 1$, we find that $\varphi _n \to 0$
and (\ref{e1}) is reduced to the quantization of the symmetric
double-well potential. In the diabatic limit $\varphi _n = - \chi _n$
and the geometric phase compensates the tunneling one. The physical
argument leading to this compensation may be easily rationalized as follows.
Indeed, at the reflection in the crossing point $ X= 0$, the trajectories
in the classically forbidden energy region are the same as those for
the tunneling region but with a phase shift $\pi $.

We focus now on the quantization rules for the over-barrier energy region.
Closely following the consideration performed above for the tunneling region, and
replacing the connection matrix (\ref{q8}) by the corresponding matrix
for the over-barrier region (see section \ref{bas})
(and making some other self-evident replacements) 
we end up after some 
algebra with the quantization rule
\begin{eqnarray}
\label{e4} &&
(1 -\exp (-2\pi \nu ))\cos (2\gamma W_L^* - \phi )\cos (\gamma W^* +
\phi ) + \exp (-2\pi \nu )\cos ^2 \left (\gamma W_L^* + \frac{\gamma W^*}{2}
\right ) = 0
\, ,
\end{eqnarray}
where $W^*$ is the action in the well formed by the upper adiabatic potential,
and $\phi = \arg \Gamma (- i \nu ) + \Im (2 \tilde \chi) $ is determined according to
(\ref{cd}).
From the Eq. (\ref{e4}) follows that the eigenstates are determined
by the parameter
\begin{eqnarray}
\label{e5} &&
B = \frac{\exp (-2\pi \nu )}{1 -\exp (-2\pi \nu )}
\, .
\end{eqnarray}
In the diabatic limit $\nu \to 0$, and therefore, $B \to 1/(2\pi \nu)$
in (\ref{e4}) the main contribution is due to the second term,
and it leads to a splitting of degenerate levels in the diabatic potentials.
Moreover since
\begin{eqnarray}
\label{e6} &&
\gamma \left (W_L^* + \frac{W^*}{2}\right ) = \pi \left (n + \frac{1}{2} \pm
2 \nu \sin \left [\gamma \left (W_L^* + \frac{W^*}{2}\right )  + \phi \right ]
\right )  
\, ,
\end{eqnarray}
the splitting increases when the Massey parameter $\nu $ increases,
and it is an oscillating function of the interaction $U_{12}$.

In the adiabatic limit, when $\nu \to \infty $, $\phi  \to 0$, and,
therefore, from (\ref{e5}) $ B \simeq \exp (-2\pi \nu )$, the main contribution
to (\ref{e4}) comes from the first term which determines the quantization rule
for the upper one-well potential and for the lower double-well potential
in the over-barrier energy region, and in this limit the parameter $B$
plays a role of the tunneling transition matrix element. For $B$ smaller
than nearest level spacings for the lower and for the upper potentials, one can find
from (\ref{e4}) two sets of quantization rules leading to two
sets of independent energy levels
\begin{eqnarray}
\label{e7} &&
\gamma W^* = \pi \left (n_1 + \frac{1}{2}\right )
\, ; \,
2\gamma W_L^*  = \pi \left ( n_2 + \frac{1}{2}
\right ) 
\, .
\end{eqnarray}
Since the eigenstate energy level displacements depend on the adiabatic coupling
$U_{12}$
the resonances can occur at certain values of this parameter, where
the independent quantization rules (\ref{e7})
are not correct any more. The widths of these resonances
are proportional to $\exp (-2\pi \nu )$ and therefore
are strongly diminished upon the Massey parameter $\nu $ increase.
This behavior is easily understood, since in the limit the wave
functions of the excited states for the lower potential are delocalized,
and their amplitudes in the localization regions for the low-energy states
of the upper potential, are very small.

\subsection{Bound initial and decay final states: 
the diabatic potentials $(1 +
X)^2/2$ and $(1/2) - X$ crossing.}
 
The second instructive example treats
the one-well and linear diabatic potentials crossing. 
It leads to the lower adiabatic decay potential and
to the upper one-well adiabatic potential. 
The quantization rules in this case correspond to the vanishing
amplitudes for the exponentially increasing solutions when $ X \to -\infty $, and besides one has to
require that no waves propagating from the region of infinite motion, i.e. at $X > 1/2$. Performing the
same as above procedure we find that in the tunneling energy region, the eigenstates are the roots of the
following equation 
\begin{eqnarray} 
\label{e10} && 
\tan (\gamma W_L^*) = - i \frac{4}{p^2(\nu )}\exp (2
\gamma W_B^*) \, , 
\end{eqnarray} 
with the same as above notation.
To proceed further it is convenient to introduce the complex action to describe the quasi - stationary
states 
\begin{eqnarray} 
\label{e11} && 
\gamma W_L^* = \pi \left (\frac {E_n}{\Omega } - i \frac{\Gamma
_n}{2\Omega }\right ) \, , 
\end{eqnarray} 
where evidently $\Omega = \partial W_L/\partial E$ does depend
on $E$. From (\ref{e11}) the real and imaginary parts of the quantized eigenstates are 
\begin{eqnarray}
\label{e12} && 
E_n = \Omega \left (n + \frac{1}{2} \right ) \, ; \, \Gamma _n = p^2(\nu )\frac{\Omega
}{2\pi }\exp (- 2\gamma W_B^*) \, . 
\end{eqnarray} 
This relation (\ref{e12}) describes the non-adiabatic
tunneling
decay of the quasi-stationary states of the lower 
adiabatic potential. The same as we already got for the
two parabolic potentials crossing (\ref{e2}), here the 
tunneling and the adiabatic factors are entering
decay rate multiplicatively. Since the decay 
rate is proportional to the square of the tunneling matrix
element, $\Gamma _n \propto p^2(\nu )$ as it should be.
 
In the over-barrier energy region the quantization rule is 
\begin{eqnarray} 
\label{e13} && 
(1 - \exp (-2\pi \nu ) \exp [-i (\gamma W_L^* - \phi )]\cos (\gamma W^* + \phi ) + 
\\ &&
\nonumber
\exp (-2\pi \nu )\exp (-i \gamma W^*/2 ) \cos \left 
(\gamma W_L^* + \frac{\gamma W^*}{2}\right ) = 0 \, ,
\end{eqnarray} 
and the actions depend on the energy $E$ as 
\begin{eqnarray} 
\label{e14} && 
\gamma W_L^* =
\pi \frac{E}{\Omega } \, ; \, 
\gamma W^* = \pi \left [ - \gamma \frac{\Omega _0(u^\# + u_{12})}{\Omega _1} + 
\frac{E}{\Omega _1}\right ] \, , 
\end{eqnarray}
where $\Omega $ and $\Omega _1$ are $E$-dependent frequencies of the
diabatic and the upper adiabatic potentials.
 
In the diabatic limit the decay rate is proportional 
to the Massey parameter $\nu $ and has a form
\begin{eqnarray} 
\label{e15} && 
\Gamma _n \simeq \frac{\Omega _0}{2} \nu \cos^2 (\gamma W^* + \phi ) \, ,
\end{eqnarray} 
and in the opposite, adiabatic, limit the decay rate is 
\begin{eqnarray} 
\label{e16} &&
\Gamma _n \simeq \Omega _0\exp (-2 \pi \nu )( 1 - 
\sin (2 \gamma W_L^*  - \phi )) \, . 
\end{eqnarray} 
In the
both limits the decay rate is the oscillating function of $U_{12}$. 

We illustrate the dependence $\Gamma
(U_{12})$ for the crossing diabatic potentials $U_1 = (1 + X)^2/2$ and $U_2 = (1/2) - X$ in the Fig. 7.
Note that while the tunneling decay rate of the low-energy states is increased monotonically with the
Massey parameter $\nu $, the decay rate of the highly excited states goes to zero in the both (diabatic
and adiabatic) limits. Besides there are certain characteristic values of $U_{12}$ when the r.h.s. of
(\ref{e15}) or (\ref{e16}) equal to zero and therefore $\Gamma _n =0$.
This seemingly paradoxical and contradicting to conventional wisdom result
can be rationalized as follows. For the case under consideration
(one well upper adiabatic and decay lower adiabatic potentials) there are always
energy levels blocked by the upper adiabatic potential. This resonance phenomenom
manifests itself wave-like particle properties omnipresent in quantum
mechanics. For the system under consideration the upper adiabatic potential
is equivalent to a resonator with a set of well - defined modes (resonances)
with high quality factors. An important feature (in distinction to a conventional
resonators where these modes occupy more or less homogeneously the whole phase space)  
is that the resonance modes are localized in its own effective cavity whose
position is given by the conditions $\Gamma _n = 0$ (\ref{e15}) or (\ref{e16}).

Quite similar one can study the more general example, describing two
non-symmetric diabatic potentials crossing
at $X = 0$ point: 
\begin{eqnarray} 
\label{e17} && 
u_1 = \frac{1}{2}(1 + X)^2 \, ; \, u_2 =
\frac{1}{2b}(X^2 - 2 b X + b) \, . 
\end{eqnarray} In a certain sense it is the generic case, and when the
parameter $b$ entering the potential (\ref{e17}) is varied from 1 to $\infty $, we recover the two
particular examples considered above, and come from two identical parabolic potentials to the case
one-well and linear diabatic potentials crossing. 
This kind of the potential was investigated
recently by two of the authors (V.B. and E.K) \cite{BK02} aiming to study crossover behavior from coherent
to incoherent tunneling upon increase of the parameter $b$, 
the larger is this parameter $b$, the larger
will be the density of final states. 
The criterion for coherent-incoherent crossover behavior found in
\cite{BK02} based on comparison of the transition matrix 
elements and the inter level spacings in the
final state. The analogous criterion should hold for the level crossing problem, 
however in the latter
case
the tunneling transition matrix elements has to be multiplied 
by the small adiabatic factor. Therefore the
coherent - incoherent tunneling crossover region moves 
to the more dense density of final states, and the
larger $U_{12}$ is the smaller will be the region for incoherent tunneling.
Quite different situation occurs for highly excited states. In the diabatic limit, the transition matrix
element is increased with the Massey parameter $\nu $, and therefore at a given $b$ value, the system
moves to more incoherent behavior. In the adiabatic limit, the transition matrix element is exponentially
small, and coherence of the inter-well transitions should be restored. However, since the matrix elements
are oscillating functions of $U_{12}$ for the intermediate range of this coupling ($U_{12}$) coherent -
incoherent tunneling rates are also non-monotonically varying functions.

\section{Intermediate energy region.}
\label{inter}

More difficult task is to derive the quantization rule in the intermediate
energy region, where all four roots of the characteristic
equation contribute into the solutions. 
One has to use the connection matrix (\ref{q99})  
computed for this region 
(see details in section \ref{mat} and appendix \ref{A}). 
It has two $2 \times 2$ blocks structure, the same as the
connection matrices for the tunneling and over-barrier regions. 
Here, we present another derivation of the same connection matrix
using the adiabatic representation. It offers a deeper insight into the mathematical
structure of the problem, and besides provides physically relevant relations between
the adiabatic and diabatic states in the intermediate energy region.
The very possibility to use the both representation is stipulated by the fact (we
have mentioned already in section \ref{mat}) that the semiclassical eigen functions
in the intermediate energy region can be represented as linear combinations
either diabatic or adiabatic functions (this adiabatic - diabatic transformation
has been discussed for quantum coherence phenomena in \cite{FS97}, see also \cite{NU84}). 

Since the adiabatic potentials have two second order turning points (the minimum of the upper,
and the maximum of the lower adiabatic potentials) the blocks of the connection matrix
in the intermediate energy region (where now unlike the matrices (\ref{q1}), (\ref{q9}) 
describing the transitions between the diabatic states, the
connection matrix
corresponds to the transitions between the adiabatic states,
and non-adiabatic perturbations induce the transitions \cite{BV02}), are characterized by
the parameters $\tilde {q}_{1 , 2}$ analogous to $q_{1 , 2}$ from (\ref{a55}) entering (\ref{q888}). 
Respectively, for the real - valued blocks
\begin{eqnarray} 
\label{b34} 
\tilde {q}_1 
= \frac{\gamma \sqrt {2u_{12}}}{4 f}(u_{12} + \alpha ) \, , 
\end{eqnarray} 
and the complex - valued blocks (
associated with the maximum of the lower adiabatic potential) 
\begin{eqnarray} 
\label{b35} 
\tilde {q}_2  =  \frac{\gamma \sqrt {2u_{12}}}{4 f}(u_{12} - \alpha )  
 \, . 
\end{eqnarray} 
We can now reap the fruits of the previous subsection consideration
efforts. First, let us note that from the relations (\ref{b11}) and (\ref{b14}) one can see that when the
energy approaches to the top of the barrier, the exponents $p^{(i)}$ and $\tilde {p}^{(i)}$ of the parabolic
cylinder functions are increased and thus, more and more deviated from the value prescribed by the Massey
parameter $\nu $. Second, increasing of $\beta _{(i)}$ upon $|\alpha |$ decreasing, decreases the values
of $|X|$ where the asymptotic smooth matching of the solutions should be performed. For $\delta \to 0$
these $|X|$ values are located deeply in the classically forbidden region, where the potentials are close
to the diabatic potentials, while for $\delta \geq 1/4$, these coordinates $|X|$ are of the order
of the quantum zero-point oscillation amplitudes, and therefore to find the solution in this region, we
have to use the adiabatic representation.

Although as it is shown in the appendix \ref{B}, the intermediate region for the
both subregions, $S^{\prime \prime }$, at $\delta < 1/4$, and $S^\prime $, at $\delta > 1/4$
can be investigated on equal footing in the frame work of the comparison equations (i.e. at the
diabatic basis) it is instructive to study the problem also in the adiabatic
representation, what is the purpose of this section. As a sub-product of this consideration
we get also the justification of the comparison equation approach.
In the adiabatic basis the intermediate subregions $S^\prime $ and $S^{\prime \prime }$ should be studied
separately.
Two simple observations give us a conjecture how to treat 
the problem in the intermediate energy
region. First of all 
the energetical ''window'' for the
intermediate subregion $S^{\prime \prime }$, where $\delta \leq 1/4$, and $|\alpha | \leq u_{12}$, 
in terms of the dimensional energy scale is determined by the rectangle around the crossing point 
\begin{eqnarray}
\label{b22} 
U_{12} \leq 2 U_{12}^* \, ; \, |U^\# - E | \leq U_{12}^* 
\, ,
\end{eqnarray}
where we define $U_{12}^* \equiv (1/2)(\hbar ^2 F^2 /m)^{1/3}$. 
By the other words the characteristic interaction energy at the intermediate region
boundaries does not depend on $U_{12}$. 
Analogously the intermediate subregion $S^\prime $ is restricted by the lines
\begin{eqnarray}
\label{b25} 
U_{12} \geq 2 U_{12}^* \, ; \, |U^\# - E| \leq U_{12}
\, . 
\end{eqnarray}
The positions of the linear turning points $|X^*|$
corresponding to the energies $U^\# \pm U_{12}^*$ do depend on the ratio $U_{12}/U_{12}^*$. These points are
located inside or outside of the interval $[-\gamma ^{-1/2} \, ; \, +\gamma ^{-1/2}]$ at $U_{12}/U_{12}^*
< 1$ and at $U_{12}/U_{12}^* > 1$, respectively. Accordingly for the both cases 
the matching conditions in
the intermediate energy region are different. 
In the former case for the asymptotic matching region the
potentials can be reasonably approximated by parabola, and therefore we should work with the Weber
equations, and for the latter case the matching are performed in the region where the potentials are
linear ones, thus the equations are reduced
to the Airy ones.

Let us discuss first the intermediate energy subregion $S^{\prime \prime } $, where $\tilde {q}_1$ and $\tilde {q}_2$ are large,
and therefore the Massey parameter, i.e. the indices of the Weber functions
are also large. 
The arguments of the Weber functions are $\propto X\sqrt \gamma $, 
and their
asymptotic expansions determine the interval where 
the matching should be done (\ref{b25}).
In what follows we will closely follow the method we borrowed from 
Olver paper \cite{OL59} (for the asymptotic expansions of the Weber
functions with large indices, see also his
monograph \cite{OL74}), which is in fact an expansion over small parameters $1/|\tilde {q}_i|$ (where $|\tilde {q}_i|$ are
the exponents (\ref{b34}), (\ref{b35})) of the fundamental Weber solutions, 
and it leads to the following asymptotic
solution at $X > 0$ 
\begin{eqnarray} 
\label{b26} 
\Psi _+^-(X) \simeq Y_+^{-1/2} (X +
Y_+)^{-\tilde {q}_1} \exp (- \gamma X Y_+) \, , \, 
\Psi _-^-(X) \simeq Y_-^{-1/2} (X + Y_-)^{i\tilde {q}_2} \exp (i \gamma X Y_-) 
\, , 
\end{eqnarray} 
where $Y_\pm =
\sqrt {u_{12}^2 \pm \alpha ^2 + f^2 X^2}$. 
Using the known relation between the
fundamental solutions of the Weber equation \cite{EM53} 
$$ D_\mu (z) = \exp(-i\pi \mu )D_\mu
(z) + \frac{\sqrt {2\pi }}{\Gamma (-\mu )}\exp \left (-i\pi \frac{\mu +1}{2}\right )D_{-\mu -1}(i z) 
\, ,
$$ 
we can find two other (complimentary to (\ref{b26}) solutions 
\begin{eqnarray} 
\label{b27} 
\Psi _+^+(X)
= Y_+^{-1/2} \left (-\sin(\pi \tilde {q}_1)(X + Y_+)^{-\tilde {q}_1} \exp (- \gamma X Y_+) 
+ \exp (-2 \chi _1) \frac{\sqrt {2\pi }}{\Gamma ((1/2) + \tilde {q}_1)}(X + Y_+)^{\tilde {q}_1} 
\exp (\gamma X Y_+)\right ) \, , 
\end{eqnarray} 
and
\begin{eqnarray} && 
\label{b28} 
\Psi _-^+(X) = \\ && 
\nonumber
Y_+^{-1/2} \left (-i\exp (-\pi \tilde {q}_2)(X + Y_-)^{i\tilde {q}_2} \exp (i\gamma X Y_-) 
+ \exp (-2 \chi _2) \frac{\sqrt {2\pi }}{\Gamma ((1/2) - i\tilde {q}_2)}(X + Y_-)^{i\tilde {q}_2} 
\exp (-i \gamma X Y_-)\right ) \, . 
\end{eqnarray} 
In the case of weak level coupling, i.e., for the intermediate energy subregion $S^{\prime \prime }$, 
the adiabatic potentials everywhere (except 
a
small neighborhood of the level crossing point) can be linearized, i.e. represented as
$\alpha \pm f|X|$, and the asymptotic solutions are reduced to a linear combination of the
following functions 
\begin{eqnarray} 
\label{b36} 
\Phi _+^\pm \propto (f|X|)^{-1/2}\exp (\pm \xi _+ + sign X)
\, , \, \Phi _-^\pm \propto (f|X|)^{-1/2}\exp (\pm \xi _- - sign X) \, , \, \xi _\pm = \frac{2}{3 f}(f |X|
\pm \alpha )^{3/2} \, ,
\end{eqnarray}
and these functions are smoothly matched with semiclassical solutions (see details in the appendix \ref{B}).                                          
As a result we can calculate finally the connection matrix $U_c^{\prime \prime }$ in the intermediate
energy region in the adiabatic basis
\begin{eqnarray} && 
\nonumber 
\hat {U}_c^{\prime \prime } = \left [ 
\begin{array}{cc} 
(\sqrt {2\pi }/\Gamma (-i \tilde {q}_2))\exp (-2\chi (i \tilde {q}_2)) & 0   \\
0 & (\Gamma (\tilde {q}_1)/\sqrt {2\pi })\exp (2\chi (\tilde {q}_1))\sin ^2(\pi \tilde {q}_1) \\
0 & \cos (\pi \tilde {q}_1)   \\
i\exp(-\pi \tilde {q}_2) & 0  
\end{array} 
\right .  
\end{eqnarray}
\begin{eqnarray}
\nonumber
\left .
\begin{array}{cc}
0 & -i \exp (-\pi \tilde {q}_2)  \\
-\cos (- \tilde {q}_1)  &  
0  \\
(\sqrt {2\pi }/\Gamma (\tilde {q}_1))\exp (-2\chi (\tilde {q}_1)) & 0 \\
0 & 2(\Gamma (-i \tilde {q}_2)/\sqrt {2 \pi })\exp (-2i\chi (\tilde {q}_2))\exp (-\pi
\tilde {q}_2)\cosh ^2(\pi \tilde {q}_2) 
\end{array} 
\right ] \, , 
\end{eqnarray} 
where the function $\chi $ is defined in (\ref{a551}), (\ref{a552}).
We see that the connection matrix in the adiabatic basis, unlike (\ref{q888}) defined in the diabatic basis,
does not provide continuous transformation into the connection matrices for the tunneling and over-barrier
energy regions ((\ref{q8}) and (\ref{q88}) correspondingly). This apparent inconsistency
is due to disregarding of adiabatic level interactions which become relevant in the
intermediate energy region. However there is a simple remedy to ensure the continuous over all energy
windows matching of the connection matrices. One has to rotate the complex plane $q$
over the angle $\varphi $ (\ref{apen10}). 
Thus luckily (as it is often the case in semiclassical
approaches) we can safely reduce the problem quite accurate to the Weber or Airy equations in the
both intermediate energy subregions, using respectively the perturbation theory 
with respect to the diabatic or adiabatic states.
The found adiabatic connection matrix could be used on the same footing as the diabatic
connection matrix (\ref{q888}), e.g., to derive the quantization rule, which for
the intermediate energy window can be  
written in
the simple and compact form as
\begin{eqnarray}
\label{e8} &&
\cos (2\gamma W_L^*) = - \exp (-\pi \tilde {q}_2)
\, .
\end{eqnarray}

It is useful to illustrate the essence of the given above general result
by simples (but yet non trivial) examples. First, let us consider two
identical parabolic potentials with their minima at $X = \pm 1$ and with
the coupling which does not depend on $X$. Since the symmetry,
the solutions of the Hamiltonian can be represented as symmetric
and antisymmetric combinations of the localized functions
\begin{eqnarray}
\label{e9} &&
\Psi ^\pm = \frac{1}{\sqrt 2}( \Phi _L \pm \Phi _R)
\, .
\end{eqnarray}
The functions are orthogonal, and, besides, two sets of the functions
$(\Psi _e^+ \, , \, \Psi _0^-)$, 
and $(\Psi _0^+ \, , \, \Psi _e^-)$ 
(where the subscripts $0$ and $e$ stand for the ground and for
the first excited states respectively) correspond to the two possible kinds of level crossings.

In Fig. 8 we depicted schematically the dependence of the
level positions on the coupling $U_{12}$. In the energy region $E \leq U^* + U_{12}$
where only there exist the discrete levels of the lower
adiabatic potentials, there are the pairs of the alternating parity
levels
$(\Psi _e^+ \, , \, \Psi _0^-)$, 
and $(\Psi _0^+ \, , \, \Psi _e^-)$. The tunneling splittings are increased   
monotonically since the Massey parameter $\nu $ is increased, and the
barrier is decreased with $U_{12}$. The same level and parity
classification is remained correct for the energy region above
the barrier of the lower adiabatic potential where the spectrum
becomes almost equidistant one. However, in the over-barrier region, the resonances
are occurred between the levels of the same parity, and this sequence
of the odd and of the even levels is broken, and level displacements
are not monotonic functions of $U_{12}$. Some of the levels of different
parities can be mutually crossed. For the upper adiabatic potential the level
sequence is opposite to this for the lower
adiabatic potential. 
The intermediate subregion $S^{\prime \prime }$ limits are shown by two dashed lines.
The boundaries between the intermediate subregion $S^\prime $ and the tunneling
and the over-barrier regions are shown by the dotted-dashed lines outgoing from the
corners of the subregion $S^{\prime \prime }$ rectangle, and these lines
coincide with energetic displacements of the top and of the bottom of the adiabatic
potentials.
Note also that we checked the results of our
semiclassical approach and found remarkably good agreement
with the numerical quantum diagonalization. Shown in the Fig. 8 
level displacements versus $U_{12}$
coincide (with the error not exceeding $10 \% $ for the full range of variation of $U_{12}$, including
the both intermediate energy subregions) with the results of the numerical
diagonalization in the basis of harmonic oscillator functions of the initial Hamiltonian (\ref{nn3} for two diabatic crossing potentials
$(1 \pm X)^2/2$.

\section{Coupling to a thermal reservoir.}
\label{bath}

We have considered semiclassical quantization of bound and quasi-stationary
states beyond the adiabatic approximation but for $1D$ case only. 
Of course the energetic profile of any real system is characterized by a
multidimensional surface. However, it is often possible to identify a reaction coordinate, such that the
energy barrier between initial and final states is minimized along this specific direction, and,
therefore, effectively one can treat the system under consideration as $1D$, regarding 
all other degrees of freedom 
as a bath of harmonic oscillators.
In this section we investigate the simplest multidimensional Hamiltonian
describing the non-adiabatic transitions, namely the $2 \times 2$ matrix potential
for the $X$ variable (or what is the same two $1D$ diabatic potentials crossing considered
in the previous sections) and the set of ''transverse'' harmonic oscillators
$\{ Y_k\}$ coupled with the reaction coordinate $X$
\begin{eqnarray}
\label{t1} &&
V(X , \{Y_k\}) = V_1 (X) + \sum _{k}\frac{\omega _k^2}{2} Y_k^2 + F(X)\sum _kC_k Y_k
\, .
\end{eqnarray}
Here $V_1(X)$ is the bare (in a general case anharmonic) $1D$ potential,
$\omega _k$ is the eigenfrequency of the transverse oscillator $k$,
the function $F(X)$ describes how the only strongly fluctuating coordinate $X$
is coupled to thermal bath of transverse oscillators, and $C_k$ are corresponding
coupling constants.
This kind (\ref{t1}) of the multidimensional potential has
been studied in the literature (see e.g. \cite{BM94}), and some
efforts were made to find a feasible approximation to treat the
potential within the semiclassical approach. In this section we legitimate
the method proposed in \cite{BM94} focusing on the LZ problem in the tunneling
region. Similar consideration can be easily generalized for the over-barrier
and intermediate regions.

The equations of classical motion (in imaginary time) for the transverse
coordinates have the following form 
\begin{eqnarray}
\label{t3} &&
\ddot X = \frac{d V_1}{d X} -  \sum _{k}\frac{C_k^2}{\omega _k^2} \frac{d F(X)}{d X} I(\omega _k,
[F(X)])
\, ,
\end{eqnarray}
where $III$ is the integral transformation
\begin{eqnarray}
\label{t4} &&
I(\omega _k, [F(X)]) = \frac{\omega _k}{2} \int _{-\infty }^{\infty } \exp (-\omega _k
|t - t^\prime |F(X (t^\prime ))d t^\prime
\, .
\end{eqnarray}
It can be expand in the following high- and low-frequency limits
\begin{eqnarray}
\label{t5} &&
I(\omega , [F]) =
\left \{ 
\begin{array}{c} 
F + \omega ^{-2}\ddot F + \omega ^{-4} \ddot {\ddot F} + .... \, , \omega \to \infty
\\ 
-\omega ^2 R_2 - \omega ^4 R_4 - ....\, , \omega \to 0
\end{array} 
\right . 
\, , 
\end{eqnarray} 
where
\begin{eqnarray}
\label{t6} &&
R_n = \int _{-\infty }^{t} dt_1 \int _{-\infty }^{t_1} dt_2....\int _{-\infty }^{t_{n-1}}F(t_n)
dt_n
\, .
\end{eqnarray}

At the high frequency limit (\ref{t5}) is reduced to the trajectory
equation but with the renormalized potential corresponding to the following 
$X$-dependent effective
mass
\begin{eqnarray}
\label{t7} &&
\sqrt {m^*} \frac{d}{d t} [\sqrt {m^*} \dot X] = \frac{d \tilde V}{d X} + O (\rho _6)
\, ,
\end{eqnarray}
where
\begin{eqnarray}
\label{t8} &&
m^*(X) = 1 + \rho _4 \left (\frac{d F}{d X}\right )^2 \, , \,
\tilde V (X) = V_1(X) - \frac{1}{2}\rho _2 \frac{d F^2}{d X}\, , \,
\rho _n \equiv \sum _k \frac{C_k^2}{\omega _k ^n}
\, 
\end{eqnarray}
(remind that we put unity the bare mass $m$ in our dimensionless units).

In the low-frequency limit the trajectory equation reads
\begin{eqnarray}
\label{t9} &&
\ddot X = \frac{d V_1}{d X} - \rho _0 R_2(t)
\, .
\end{eqnarray}
In the $\rho _4$ approximation for the 
spectral density
of oscillators the neglected in (\ref{t8}) terms proportional to $\rho _6$
(and the last term in
(\ref{t9}))
are small.
The physical message of the calculation performed in this section is
that the renormalization of the effective mass leads to slowing down
of the motion and it is equivalent to say that the Massey
parameter is renormalized
\begin{eqnarray}
\label{t10} &&
\nu \to \nu ^* = \nu \sqrt {m^*(X_c)}
\, ,
\, 
\end{eqnarray}
where $X_c$ is the
crossing point.
Of course the coupling will also change the action along the
extremal action trajectory (this effect has been discussed in the literature,
see e.g. \cite{BM94}). The specific for the LZ problem new phenomenom
is renormalization of the Massey parameter (\ref{t10}) which controls
main features of the behavior for any system undergoing level crossing.

\section{Conclusion.}
\label{discus}

In conclusion we end up stressing again the main point of our methodology.
We have shown that the comparison equations for the 4-th order
differential Landau-Zener equations in the coordinate space
can be represented as two decoupled Weber equations. The indices
and the arguments of the corresponding Weber functions defined by the roots
of the characteristic equation (\ref{b8}) for the complex wave vector $\kappa $,
and $|\kappa | \gg 1$ in the semiclassical approximation.
In the frame work of our method the diabatic potential crossing points
are treated as two second order turning points characterizing by different
Stokes constants \cite{HE62}. The accuracy of the method depends
on anharmonic terms, which are not taken into account in the comparison equations,
but which are small in the semiclassical approach 
over small parameters $\delta $, $\tilde \delta $, or $\delta _{int}$
respectively, 
in the tunneling, 
over-barrier
and intermediate subregion $S^{\prime \prime }$ energy windows.
In the subregion $S^\prime $, $\delta _{int}$ is not a small parameter.
However, since the asymptotically smooth matching is performed
at small $|X| < \gamma ^{-1/2}$, anharmonic corrections to the comparison equations can be safely
neglected for this subregion as well.

We have presented detailed semiclassical analysis of
crossing diabatic potentials problem.
We examine one important (and overlooked in all previous investigations)
aspect of well - known energy level quantization problem for crossing
diabatic potentials. We derive 
the semiclassical quantization rules for the particular
situation of crossing diabatic potentials with localized initial and localized or delocalized
final
states, in the intermediate energy region, when all four adiabatic states
are coupled and should be taken into account. In fact it 
exhausts all cases practically relevant for spectroscopy of non-rigid molecules
(i.e. with more than one stable configuration).

We use the connection matrix methodology which presents
a simple and standardized description of any semiclassical approximation,
which offers therefore a deeper insight into the mathematical
and physical structure of the approximation.
We found that 
in the tunneling region the tunneling splitting is represented as a product of the
splitting in the adiabatic potential and the non-trivial function $p(\nu )$ (we calculated 
analytically) depending on the
Massey parameter, i.e. on the energy and the slopes of the diabatic
potentials in the crossing
region. 
In the over-barrier
region we found specific resonances between the levels
in the lower and in the upper adiabatic potentials and in that condition
one may not use independent quantization rules. 
New results have emanated from our consideration of the
intermediate energy region.
For this energy region we calculated the energy level quantization, 
using adiabatic basis.

We have presented in this paper all details of the LZ problem for two electronic states 
using
the connection matrix approach for the LZ problem in the coordinate space, the                   
approach which turned out very efficient for this class of problems, and are important in
many areas of pure and applied sciences.
Even though only model potentials are investigated here, our approach 
is 
quite general and
has potential applicability for various systems in physics and chemistry, 
and the results
can be tested
by their experimental consequences for many examples of molecular systems  
undergoing conversion of electronic states, non-radiative transitions, or
isomerization reactions, and not only. The results of the LZ-problem investigations
are very relevant for slow atomic or molecular collisions , \cite{MI76}, \cite{NU84},
where the interaction of diabatic potentials induces transitions between initial and
final electronic states. However, since the interaction is essential only near
the crossing point, one can compute the transition probability, linearizing
the both diabatic potentials (see our consideration in section \ref{bas}). 
The same approximation works
quite well
for the so-called pre-dissociation phenomena.

However (in contrast to the atomic and molecular collision problems) 
there are fundamental problems
of chemical physics and molecular spectroscopy
where one may not restrict oneself to the only transition probability calculations,
but should know the complete eigenvalues/eigenfunctions solution.
It is the case for example if we are interested in 
the calculation of vibrational - tunneling spectra of non-rigid molecules,
or reactive complexes with more than one stable configuration. The lowest
multi - well potential of such systems is formed from one well diabatic potentials
crossing corresponding to each stable configuration. Apart from the lowest potential,
the upper adiabatic potential with its minimum above the maximum of the lowest
potential should be also taken into account for these situations (see Fig. 1).
In the most of the calculations of tunneling splittings in the ground and low excited vibrational
states the coupling to the upper potential are neglected, what is certainly correct
only for strong enough adiabatic coupling. Evidently it is not
the case for the levels close to the adiabatic barrier top, and especially in the
upper potential well. The quantization of these levels play noticeable in the spectroscopy
of non-rigid molecules, and the same situation takes place for systems undergoing
the Jahn - Teller effect, where the interference of the diabatic states occurs in this
energy region \cite{BE84}.

One more example for the application of our results is molecular 
radiationless transitions
within excited electronic states. 
Typically for this situation the decay potential is formed owing
to crossing of bound and unbound diabatic potentials. Since
the radiationless transitions are followed by luminescence and chemical reaction
phenomena (see e.g. \cite{FR80} - \cite{ZT01}) one should know the complex 
eigenvalues of the
quasistationary states prepared by optical pumping.

Let us also stress that in real systems the characteristic values
of the coupling between the diabatic states can vary within
the very wide range from several $eV$ for the electronic
states of the same symmetry to zero (for the states with different spins).
To treat all these cases one should know the solution of the diabatic
potentials crossing problem described in our paper for the
corresponding wide range of the Massey parameter from $\nu = 0$ to $\nu \gg 1$.

\acknowledgements 
The research described in this publication was made possible in part by RFFR Grants. 
One of us (E.K.) is indebted to INTAS Grant (under No. 01-0105) for partial support.                            
The authors thank also their manuscript referees for insightful criticism.

\appendix

\section{}

\label{A}

Putting all given in the section \ref{bas} expressions (\ref{q1}) - (\ref{q33}) together we can
recapitulate the matrix elements $m_{ij}$ of the full connection matrix in the tunneling region
\begin{eqnarray} && 
\label{ap1} 
m_{11} = \frac{p}{4} \exp (-\gamma W_B^*)\cos (\gamma W_L^*)\cos (\gamma
W_R^*) - \frac{\sin ^2(\pi \nu )}{p}\exp (\gamma W_B^*)\sin (\gamma W_L^*)\sin (\gamma W_R^*) \, ; 
\\ &&
m_{12} = \frac{p}{2} \exp (-\gamma W_B^*)\sin (\gamma W_L^*)\cos (\gamma W_R^*) + 2\frac{\sin ^2(\pi \nu
)}{p}\exp (\gamma W_B^*)\cos (\gamma W_L^*)\sin (\gamma W_R^*) \, ; 
\\ && m_{21} = -\frac{p}{2} \exp
(-\gamma W_B^*)\cos (\gamma W_L^*)\sin (\gamma W_R^*) - 2\frac{\sin ^2(\pi \nu )}{p}\exp (\gamma
W_B^*)\sin (\gamma W_L^*)\cos (\gamma W_R^*) \, ; 
\\ && 
\label{ap11} m_{22} = -p\exp (-\gamma W_B^*)\sin
(\gamma W_L^*)\sin (\gamma W_R^*) + 4\frac{\sin ^2(\pi \nu )}{p}\exp (\gamma W_B^*)\cos (\gamma W_L^*)\cos
(\gamma W_R^*) \, ; 
\\ && 
m_{13/24} = \pm \cos (\pi \nu )\exp (\pm \gamma W_B^*/2)\sin (\gamma W_R^*) \, ;
\, m_{14} = -\frac{1}{2} \cos (\pi \nu ) \exp (-\gamma W_B^*/2)\cos (\gamma W_R^*) \, ; 
\\ &&
\label{ap111} 
m_{23} = -2 \cos (\pi \nu )\exp (\gamma W_B^*/2)\cos (\gamma W_R^*) \, ; \, m_{31/42} = \pm
\cos (\pi \nu )\exp (\pm \gamma W_B^*/2)\sin (\gamma W_L^*) \, ; 
\\ && 
\label{ap1111} 
m_{41} =
\frac{1}{2}\cos (\pi \nu )\exp (-\gamma W_B^*/2)\cos (\gamma W_L^*) \, ; \, m_{32} = 2\cos (\pi \nu )\exp
(\gamma W_B^*/2)\cos (\gamma W_L^*) \, ; 
\\ && 
\label{ap11111} 
m_{33} = p \, ; \, m_{44} = \frac{\sin
^2(\pi \nu )}{p} \, ; \, m_{34} = m_{43} = 0 \, . 
\end{eqnarray} 
For the over-barrier region the full
connection matrix could be given in a more compact form. 
Using (\ref{q9}), (\ref{q88}), (\ref{q333}),
(\ref{q2}) from the main body of the paper we get the following matrix 
\begin{eqnarray} && 
\label{ap2}
\left [ 
\begin{array}{cccc} 
(s/2)\cos (\gamma W_{LR} - \phi ) & s\sin (\gamma W_{LR} - \phi ) & -\exp
(-\pi \nu )\sin (\gamma W_{R\, *}) & -\exp (-\pi \nu )/2\cos (\gamma W_{R\, *}) \\ 
-s\sin (\gamma W_{LR} -
\phi ) & 2s\cos (\gamma W_{LR} - \phi ) & -2\exp (-\pi \nu )\cos (\gamma W_{R\, *}) & \exp (-\pi \nu )\sin
(\gamma W_{R\, *}) \\ 
-\exp (-\pi \nu )\sin (\gamma W_{L\, *})  & 2\exp (-\pi \nu )\cos (\gamma W_{L\, *})
& 2s\cos (\gamma W^* + \phi ) & -s \sin (\gamma W^* + \phi ) \\ 
\exp (-\pi \nu )/2\cos (\gamma W_{L\, *})
& \exp (-\pi \nu )\sin (\gamma W_{L \, *}) & s\sin (\gamma W^* + \phi ) & (s/2)\cos (\gamma W^* + \phi )
\end{array} 
\right ] \, , 
\end{eqnarray} 
where $W_{LR} \equiv W_L^* + W_R^*$, and $W_{L/R, *} \equiv
W_{L/R}^* + W^*/2 $.

\section{}
\label{B}
The efficency of the standard instanton approach \cite{PO77}, \cite{CO85} (see also \cite{BM94}, \cite{BV03})
is based on a successful choice of the comparison equation near second order turning points, where
asymptotically smooth matching of semiclassical solutions to the solutions of this equation
should be performed. It is known for example \cite{BV02} that for anharmonic potentials the Weber
equation provides  such a very successful choice since in the matching region anharmonic corrections
are still small. The aim of this appendix is to show that the analogous situation holds for
crossing diabatic potentials points, where two Weber equations can be successfully used as the comparison equations to
the fourth order Landau-Zener equation (\ref{nn4}). The arguments and the indices of the fundamental solutions 
to these Weber comparison equations are determined by the roots of the corresponding
characteristic equations (see below and the main body of the paper).

To prove the statement let us first substitute (\ref{nn11}) into the equation (\ref{nn7}). We get
\begin{eqnarray} 
\label{apen1} 
D^4\Phi + 4\kappa D^3 \Phi + (6 \kappa ^2 - 2 \alpha \gamma ^2)D^2 \Phi
+ 4(\kappa ^3 - \alpha \gamma ^2 \kappa  - \frac{1}{2}\gamma ^2 f) D\Phi + 
\end{eqnarray}
\begin{eqnarray}
\nonumber
[\kappa ^4
- 2\alpha \gamma ^2 \kappa ^2 - 2 \gamma ^2 f \kappa + \gamma ^4 (\alpha ^2 - u_{12}^2 - f^2X^2)]\Phi = 0 
\, ,
\end{eqnarray}                                                                                                  
where $ D^n \equiv d^n/d X^n$.
The equation (\ref{apen1}) can be formally derived by simple manipulations (two sequential differentiations
and summations) from the following second order equation
\begin{eqnarray} 
\label{apen2} 
D^2\Phi + (a_0 + a_1 X + a_2 X^2)\Phi = 0 
\, ,
\end{eqnarray}                                                                                                  
where the coefficients are
\begin{eqnarray} 
\label{apen3} 
a_0 = \kappa ^2 - \alpha \gamma ^2 - \frac{\gamma ^2 f}{2 \kappa }(1 + \delta )\, ;
\, a_1 = \gamma ^2f \delta \, ; \, a_2 = -\gamma^2 f\kappa \delta 
\, ,
\end{eqnarray}                                                                                                  
where $\kappa $ should be found from the characteristic equation (\ref{b8}), 
and $\delta $ is given by (\ref{b9}).

The fundamental solutions to (\ref{apen2}) read as
\begin{eqnarray} 
\label{apen4} 
D_p\left [\pm \left (\frac{\gamma ^4 f^2}{\kappa ^2}\right )^{1/4} \left (X - \frac{1}{2\kappa }\right )\right ]
\, ,
\end{eqnarray}                                                                                                  
where
\begin{eqnarray} 
\label{apen5} 
p = -\frac{1}{2} + \left (\frac{\gamma ^4 f^2}{\kappa ^2}\right )^{-1/2} \left (a_0  - \frac{a_1^2}{4 a_2}\right ) 
\, .
\end{eqnarray}                                                                                                  
In the tunneling (\ref{j1}) and over-barrier (\ref{j2}) regions of energies 
these 4 solutions (2 solutions of (\ref{apen4}) for two largest modulus roots of
the characteristic equation (\ref{b8})) can be separated into two independent pairs.
In the tunneling region the two largest modulus roots of (\ref{b8}) are
(two other roots are small and do not satisfy semiclassical approach)
\begin{eqnarray} 
\label{apen6} 
\kappa = \kappa _0\left (1 \pm \frac{\delta ^2}{2}\frac{\kappa _0^2}{2\kappa _0^2 - \alpha \gamma ^2}\right )
\, ; \,
\kappa _0 = \frac{\gamma }{\sqrt 2}\left (\alpha + \sqrt {\alpha ^2 - u_{12}^2}\right )^{1/2} 
\, .
\end{eqnarray}                                                                                                  
Putting (\ref{apen6}) into (\ref{apen5}) we find (neglecting $\delta ^2$ terms, i.e., for $\kappa = \kappa _0$)
4 fundamental solutions to the 
comparison equation in the form (\ref{nn13}).  
Thus from the given above expressions and (\ref{b9}), (\ref{b10}) from the main text
we conclude that the solutions $\Theta _{L/R}$ (\ref{nn13}) can be expanded over our
small parameter $\delta $, and due to the condition (\ref{b9}) anharmonic corrections
to the Weber functions (\ref{apen4}) are small (by other words the parameter $\delta $ determines the accuracy 
of our approximation).
Indeed the anharmonic terms neglected in the Weber comparison equations are of the order of $\delta $
(it is an upper estimation at $X = \alpha /f$, i.e. at the boundaries of the intermediate energy
region), thus the corrections are small according to (\ref{b9}). 
The same kind of analysis can be performed in the over-barrier region (\ref{j2}), where one finds
two imaginary largest modulus roots of the characteristic equation.
The roots are given by (\ref{apen6}) with
$\kappa _0$ and the small parameter $\tilde \delta $ defined according to (\ref{b12}), (\ref{b13}).

One simple observation helps to perform the same analysis for the intermediate energy region (\ref{j3}).
Indeed, since the differences between the
solutions to the characteristic equations for $\lambda $ (\ref{b2})
and for $\kappa $ (\ref{b8}) determine the accuracy of our approach,
let us compare the solutions. The roots of (\ref{b2}) at $X=0$ 
\begin{eqnarray} 
\label{apen7} 
\lambda _{1 , 2} \simeq \pm \gamma \sqrt {\alpha + u_{12}} 
\, ; \,
\lambda _{3 , 4} \simeq \pm \gamma \sqrt {\alpha - u_{12}} 
\end{eqnarray}                                                                                                  
are moved upon the variation of $\alpha $ in the intermediate energy region
from the real to imaginary coordinate axis. Analogously the roots of (\ref{b8})
\begin{eqnarray} 
\label{apen8} 
\kappa _{1 , 2} \simeq \pm \frac{\gamma }{\sqrt 2} \left (\alpha + \sqrt {\alpha ^2 - u_{12}^2}\right )^{1/2} 
\, ; \,
\kappa _{3 , 4} \simeq \pm \frac{\gamma }{\sqrt 2} \left (\alpha - \sqrt {\alpha ^2- u_{12}^2}\right )^{1/2} 
\end{eqnarray}                                                                                                  
are moved along the real and imaginary axis in the tunneling and in the over-barrier regions respectively.

We conclude from the (\ref{apen7}) and (\ref{apen8}) that in the tunneling and in the over-barrier
energy regions there is one-per-one correspondence between the roots $\lambda $ of (\ref{b2})
and $\kappa $ of (\ref{b8}). Just this correspondence allows us to match smoothly the semiclassical
solutions to the Schr\"odinger equation and the Weber functions found as the solutions to the
comparison equations. It is not the case in the intermediate energy region where two roots of (\ref{b2})
are real and two are imaginary ones having the same modulus, i.e. moving upon $\alpha $ variation along a circle
with the radius $\gamma \sqrt {u_{12}/2}$. In this case the semiclassical solutions can be presented as certain
linear combinations of the comparison equation solutions. We have found these combinations in the
adiabatic basis in the section \ref{inter}.
In this appendix we show how to solve the same problem in the diabatic basis, and it reveals
more clearly and explicitely an estimate of the omitted terms in the equation and
the areas where the solutions become wrong and where the matching procedure is carried out.
Indeed the roots of (\ref{b8}) in the intermediate energy region (\ref{j3}) are
\begin{eqnarray} 
\label{apen9} 
\kappa _{1 , 2} \simeq \pm \gamma \sqrt {\frac{u_{12}}{2}}\exp (i\varphi ) 
\, ; \,
\kappa _{3 , 4} \simeq \pm i \gamma \sqrt {\frac{u_{12}}{2}}\exp (-i\varphi )
\, , 
\end{eqnarray}                                                                                                  
where
\begin{eqnarray} 
\label{apen10} 
\tan \varphi = \sqrt {\frac{u_{12}- \alpha }{u_{12} + \alpha }}
\, . 
\end{eqnarray}                                                                                                  
Correspondingly to these roots (\ref{apen10}) the arguments and the indices
of the Weber functions (\ref{apen4}), (\ref{apen5}) read as
\begin{eqnarray} 
\label{apen11} 
z_1 = z_2 = 2 \kappa _{int}\sqrt {\delta _{int}} \exp (- i \varphi /2)(X + (2 \kappa _{int})^{-1}\exp (-i \varphi )
\, ; \,
\end{eqnarray}
\begin{eqnarray}
\nonumber
z_3 = z_4 = 2 \kappa _{int}\sqrt {\delta _{int}} \exp (i \varphi /2)(X + (2 \kappa _{int})^{-1}\exp (i \varphi )
\, , 
\end{eqnarray}                                                                                                  
and
\begin{eqnarray} 
\label{apen12} 
p_1 = p_2 - 1 = -1 - \frac{1}{4 \delta _{int}} \exp (- i \varphi )(1 + \delta _{int}^2 \exp (-2 i \varphi ))
\, ;\,
\end{eqnarray}
\begin{eqnarray}
\nonumber 
p_4 = p_3 - 1 = -1 - \frac{1}{4 \delta _{int}} \exp (i \varphi )(1 + \delta _{int}^2 \exp (2 i \varphi ))
\, , 
\end{eqnarray}                                                                                                  
where $\kappa _{int} = \gamma (u_{12}/2)^{1/2}$, and $\delta _{int} = (\gamma ^2 f)/(4 \kappa _{int}^3)$.

Using known due to Olver (\cite{OL59}, \cite{OL74})
asymptotics of the Weber functions we are in the position to compare the semiclassical functions
with the solutions to the comparison equations. The former functions determine by the exponential factor
\begin{eqnarray} 
\label{apen13} 
F_0^{\pm }(X) = \gamma \sqrt {u_{12} \pm \alpha } X + \frac{\gamma f^2}{12 u_{12} \sqrt {u_{12} \pm \alpha }} X^3
\, , 
\end{eqnarray}                                                                                                  
while the exponential factors entering corresponding asymptotics of the Weber functions are
\begin{eqnarray} 
\label{apen14} 
F_{1 , 2}(X) = \gamma \sqrt {u_{12} \pm \alpha }(1 + \delta _{int}) X 
\pm \kappa _{int}^2 \delta _{int}^2\exp (-2 i \varphi ) X^2 + \frac{\gamma f^2}{12 u_{12} \sqrt {u_{12} \pm \alpha }}
\left [1 \pm \frac{\alpha }{u_{12}} - \delta _{int}\right ] X^3
\, .
\end{eqnarray}                  
Let us consider now the intermediate subregion $S^\prime $, $|\alpha | \leq (f/\gamma )^{2/3}$, and $u_{12} \leq 2/\gamma $,
(see (\ref{b22})), where (\ref{b9}) does not hold. Luckily, however, the asymptotically smooth matching is
performed at small $|X| < \gamma ^{-1/2}$, where the comparison equation (\ref{apen2}), and, therefore,  
the characteristic equation (\ref{b8}) are valid (although, $\delta $ is not a small parameter). In this subregion
we have to take into consideration the term $R(\kappa , \delta )$ in (\ref{b8}). At $\alpha = 0$, and $u_{12} =0$,
the characteristic equation has one double degenerate root $\kappa = 0$, or correspondingly in (\ref{apen2}), $a_2=0$.
Thus the comparison equations are reduced to two decoupled Airy equations.
Using known Olver asymptotics for the Weber functions with
large arguments and indices \cite{OL59}, \cite{OL74}
\begin{eqnarray} 
\label{apen15} 
D_p(z) \propto \exp \left [ - \frac{1}{2} \int \left (z^2 - 4\left (p + \frac{1}{2}\right )\right )^{1/2} dz \right ]
\, 
\end{eqnarray}                  
we can find asymptotics to the solutions of (\ref{apen2})
\begin{eqnarray} 
\label{apen16} 
\Phi _0 \propto \exp \left (-i \int \sqrt {a_0 + a_1X + a_2X^2} dx \right )
\, 
\end{eqnarray}                  
valid at arbitrary values of the parameters $a_i$ ($a_2=0$ including).
This relation (\ref{apen16}) provides asymptotically smooth matching of the semiclassical solutions
with the Weber functions in the intermediate subregion $S^{\prime \prime }$ (where $\kappa $ is of the order
of $\gamma \gg 1$, and with the Airy
solutions in the subregion $S^\prime $, when $\kappa \simeq \sqrt \gamma $.
                                                                                
This consideration provides the justification of our approach described in the main body of the paper.
As it is seen from (\ref{apen13}), and (\ref{apen14}) at small $\alpha $ the accuracy of
the asymptotically smooth matching of the semiclassical solutions with the Weber functions
is of the order of $\delta _{int}$, and close to the energetic boundaries (\ref{j3}) of the
intermediate region, anharmonic corrections ($X^3$) are increased.
Thus we conclude that the matching for this case (\ref{j3}) can be performed either in the
adiabatic basis (as it has been done in the section \ref{inter}) or in the diabatic basis as
we have shown in this appendix. 
The simplest way to prove the equivalence of the both representation is to transform into
exponential forms the factors like $(X + Y_+)^{q_1}$ etc, entering the solutions (\ref{b27}), (\ref{b28}),
found in the section \ref{inter}.
In the both methods the accuracy is of the order of $\delta _{int}$,
and the connection matrices presented in the appendix \ref{A} do not depend on the basis.

\newpage

\centerline{Figure Captions.}

Fig. 1

Potentials in the vicinity of the diabatic potentials
crossing point $U^{\# }$:

The diabatic potentials (thin lines, 1,2), the adiabatic potentials
(bold lines, 3,4) by bold solid lines,
the adiabatic coupling energy $U_{12}$, and
$E _0$ is the characteristic zero-point oscillations energy
in the parabolic barrier approximated the lower adiabatic potential
near its top. The tunneling energy $E$ region is shown by a
broken line.

Fig. 2

The diabatic level crossing phenomena

(a) bound initial and final states;
 
(b) bound initial and decay final states. 

Fig. 3

Connection matrices for the tunneling energy region:

(a) in the WKB approach to the lower states, where $M^{\pm }$ are the connection matrices
for the linear turning points, and $U_c$ for the crossing point;
the shift matrices are depicted as arrows, in the classically accessible regions 
$L_L$ and $L_R$, and in the classically forbidden region $F_L$ and $F_R$ (for the
upper states no real-valued turning points);

(b) in the instanton type method one has two connection matrices $M_{L/R}^{(2)}$ for the
second order turning points and shift matrices $F_L$ and $F_R$ in the classically
forbidden region.

Fig. 4

Connection matrices for the over-barrier energy region.
The shift matrices from the crossing point to the inner turning points
are designated by $L$ (all other notations are the same as in the Fig.3).

Fig. 5

Connection matrices for the intermediate energy region (like 
in the tunneling region no real valued turning points for the upper states).

Fig.6

The matching of the asymptotic solutions in the tunneling region
for the diabatic levels crossing shown in Fig. 1a:

1 - the function $\Phi _L^+(X) \sqrt {2\pi }/\Gamma (1 + \nu )$;

2- the function $\Phi _L^-(X)$;

3 - the function $\Phi _R^+(X) \sqrt {2\pi }/\Gamma (1 + \nu )$;

4 - the function $\Phi _R^-(X)$;

$1^\prime $ - the function $\exp (k_0 X)D_{-1 - \nu }(\beta X)$;

$2^\prime $ - the function $\exp (k_0 X)D_{-1 - \nu }(-\beta X)$;

$3^\prime $ - the function $\exp (-k_0 X)D_{-1 - \nu }(\beta X)$;

$4^\prime $ - the function $\exp (-k_0 X)D_{-1 - \nu }(-\beta X)$.

Fig.7

$\Gamma _n$ versus $U_{12}$ for the quasi stationary states at the diabatic potentials $(1 + X)^2/2$ and
$(1/2) - X$ crossing; (a) 1 - 4 are the level energies 0.042 , 0.125 , 0.208 , and 0.292 for the lower
adiabatic potential; (b) $1^\prime - 3^\prime $ are the level energies 0.625 ; 0.708 ; 0792
for the upper adiabatic potential.

Fig.8
 
Level displacements versus $U_{12}$
for two diabatic crossing potentials
$(1 \pm X)^2/2$. Dashed lines show the intermediate
energy region (the subregion $S^{\prime \prime }$ is between the dashed lines, while
the subregions $S^\prime $ are confined to the left
pockets between the dashed and dotted-dashed lines); 
dotted - dashed lines show also displacements for the
top
and for the bottom of the adiabatic potentials. $k$, $n$, and $n^\prime $
are quantum numbers for the diabatic, and lower and upper adiabatic potentials.
Note
that shown in the figure  
level displacements coincide with the error not exceeding $10 \% $  with the results of the numerical
diagonalization in the basis of harmonic oscillator functions.

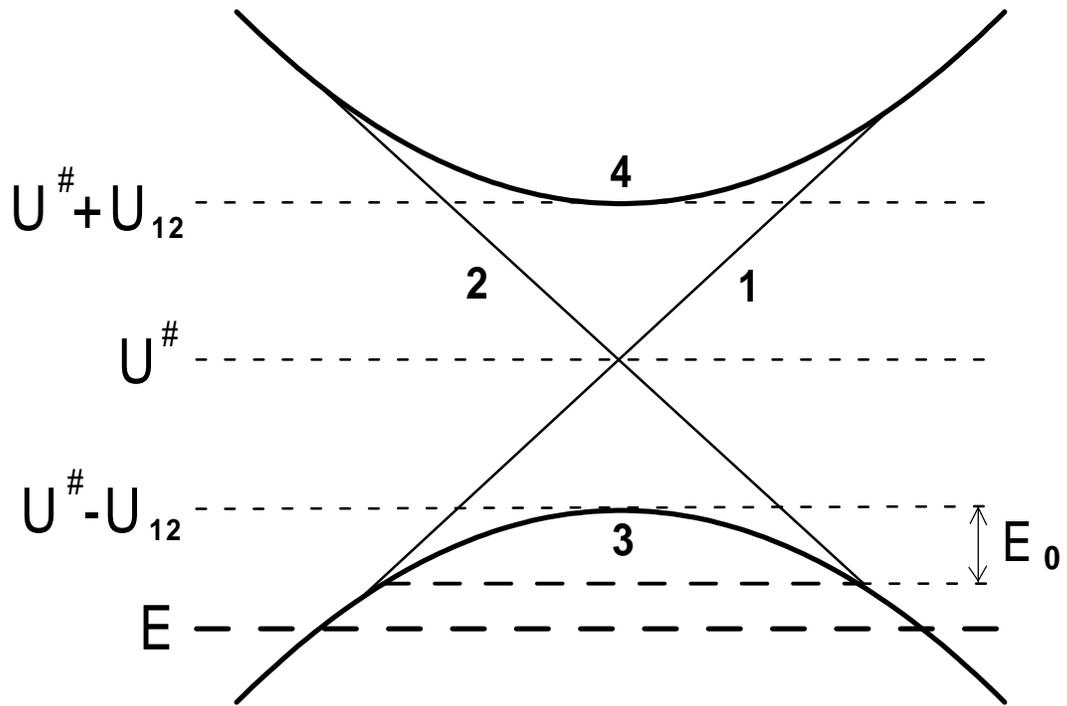
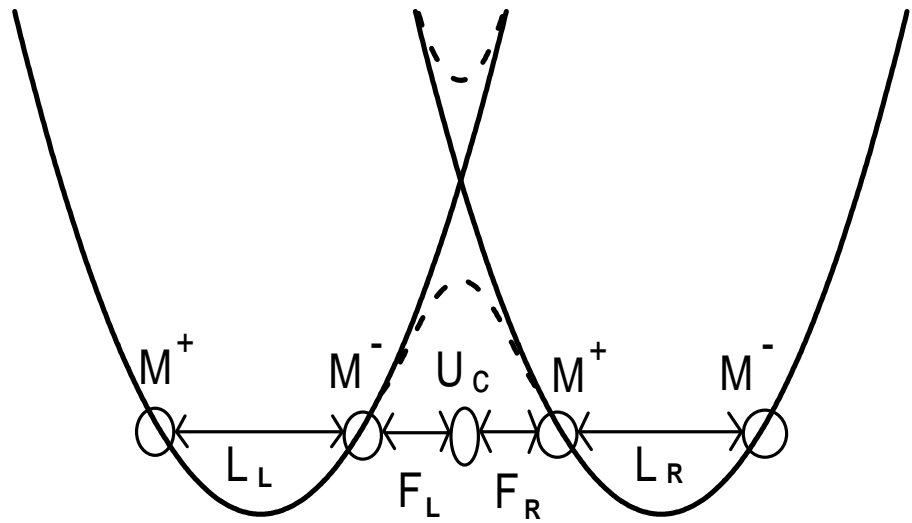
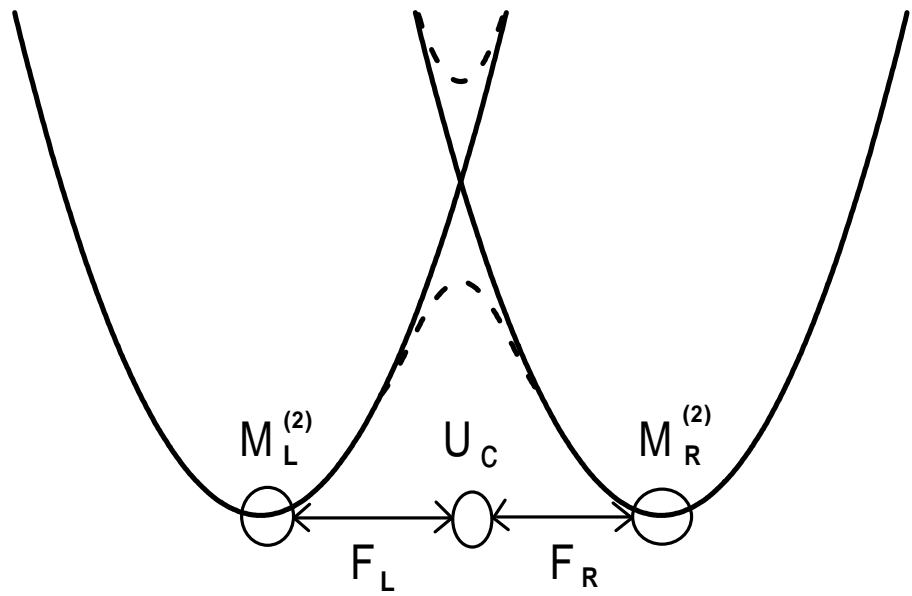
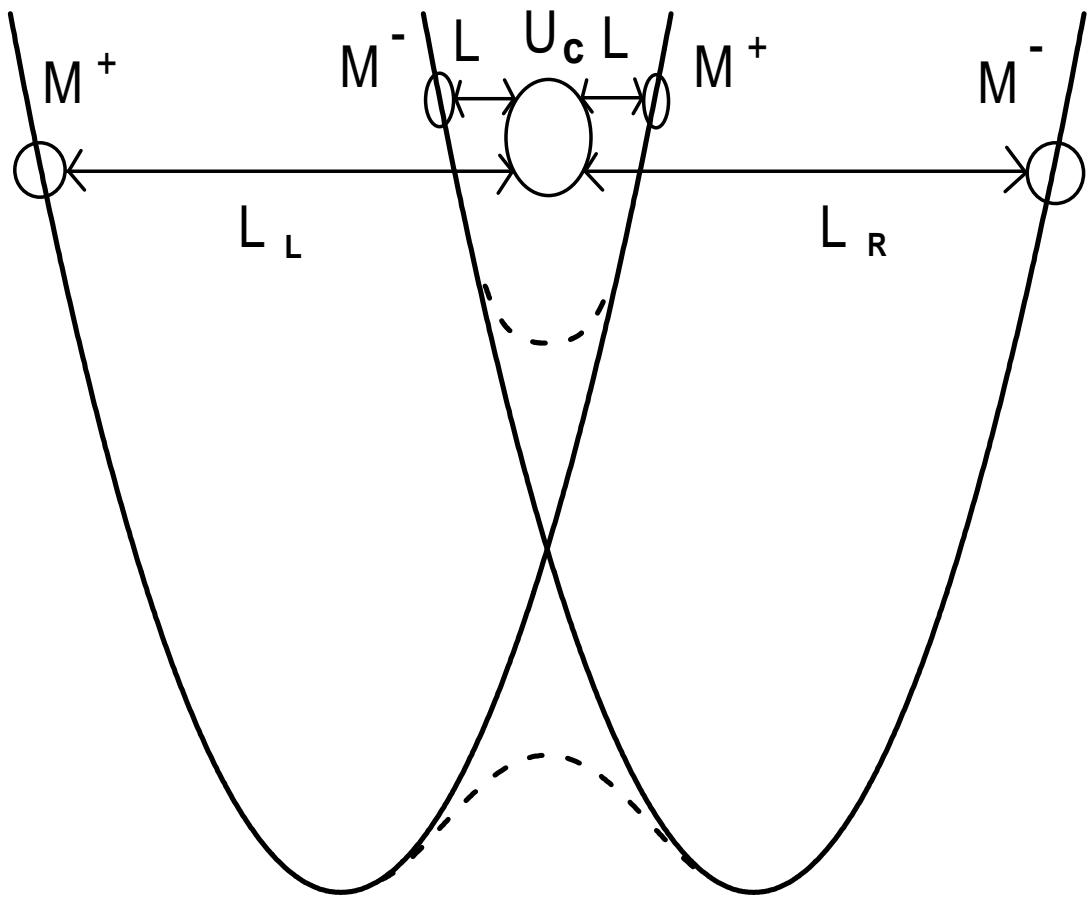
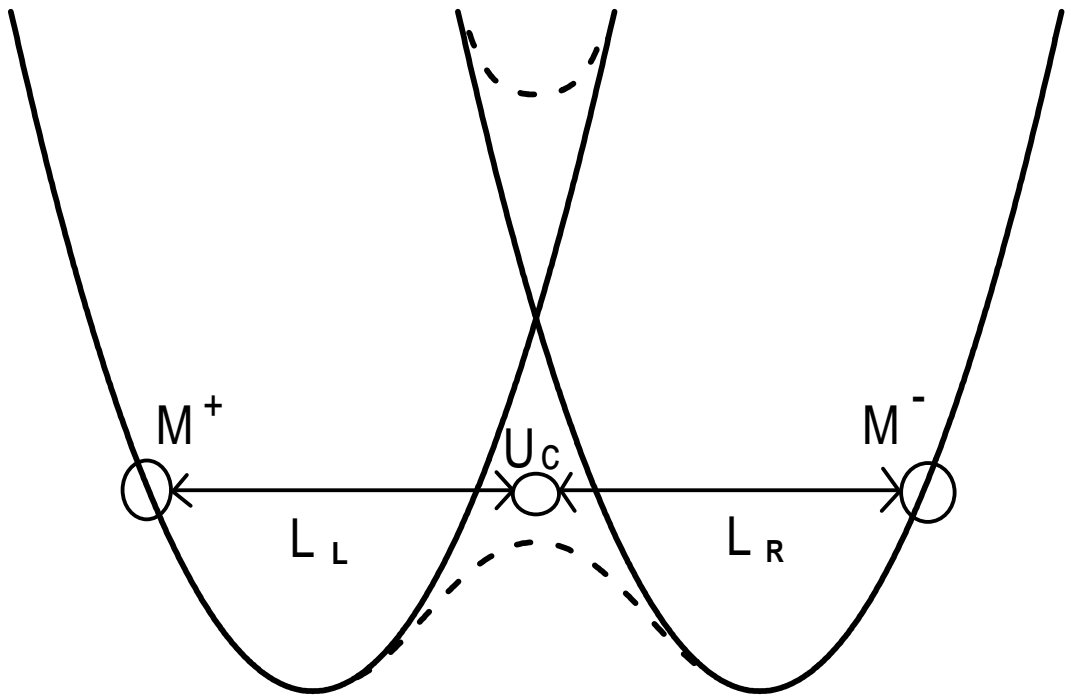
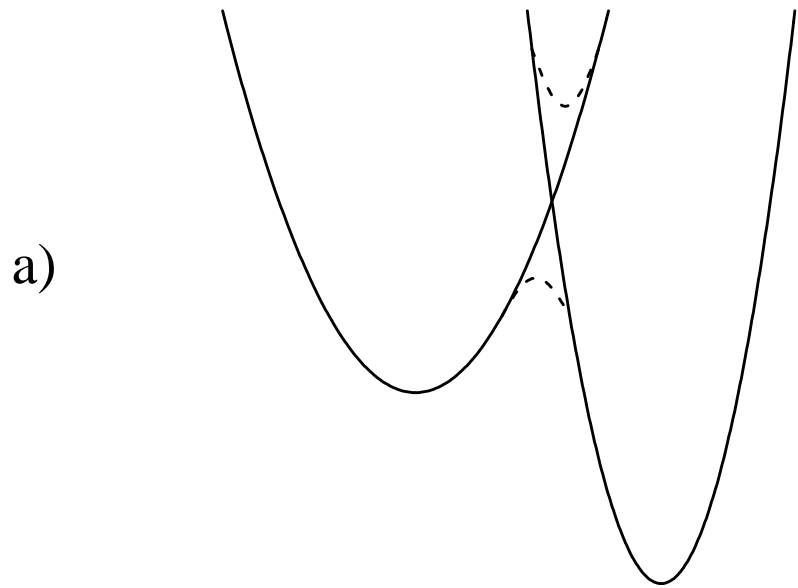
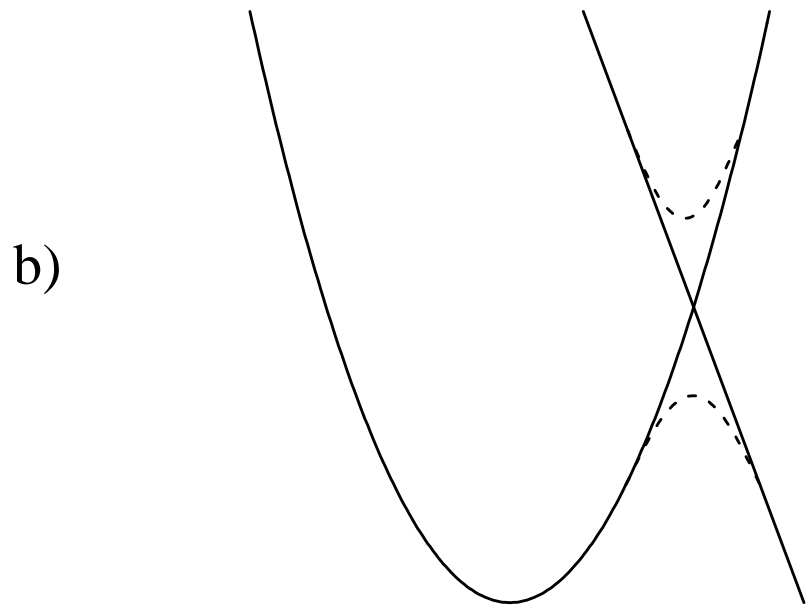
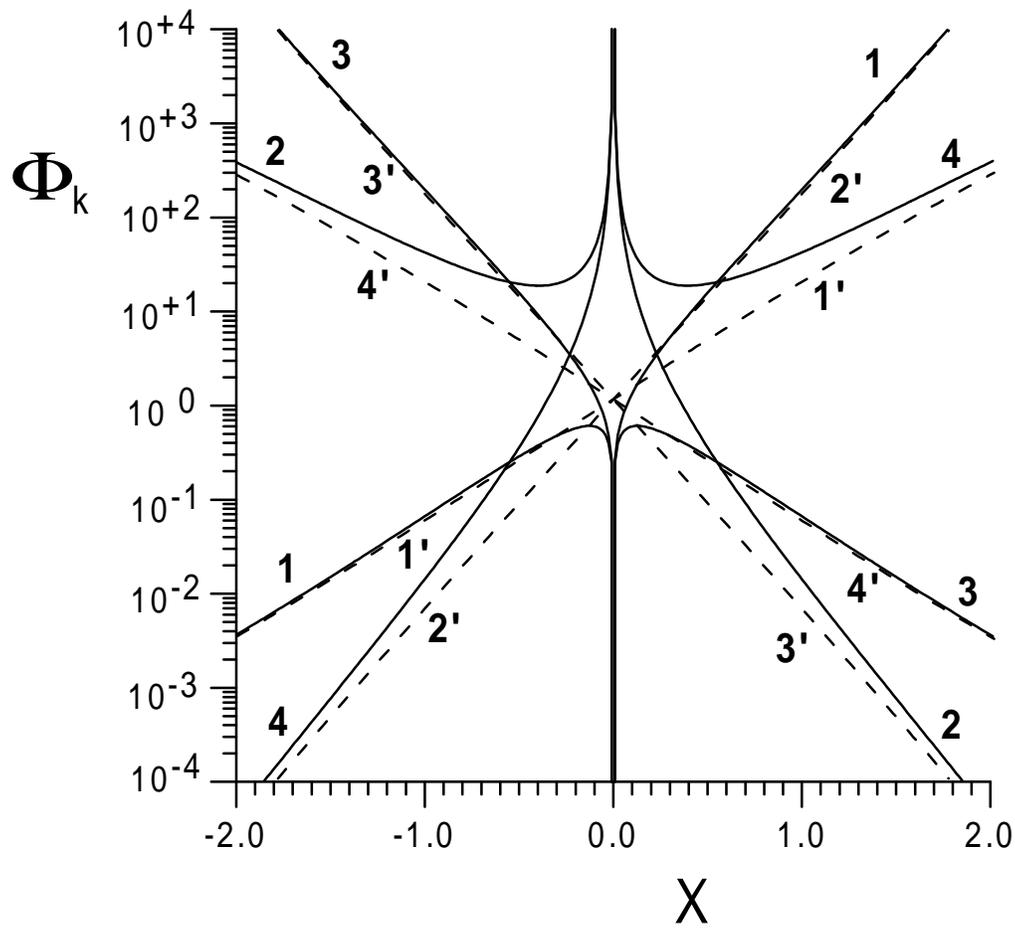
\end{document}